\documentclass[showpacs,aps,prstab,amsmath,amssymb]{revtex4-1}
\usepackage[latin9]{inputenc}
\usepackage{textcomp}
\usepackage{amsmath}
\usepackage{graphicx}
\usepackage{esint}

\makeatletter

\InputIfFileExists{t2aenc.def}{}{%
  \errmessage{File `t2aenc.def' not found: Cyrillic script not supported}}
\DeclareRobustCommand{\cyrtext}{%
  \fontencoding{T2A}\selectfont\def\encodingdefault{T2A}}
\DeclareRobustCommand{\textcyr}[1]{\leavevmode{\cyrtext #1}}

\DeclareTextSymbolDefault{\textquotedbl}{T1}

\usepackage{bm}
\usepackage{booktabs}
\usepackage[usenames]{color}
\usepackage{hyphenat}

\renewcommand{\[}{\begin{equation}}
\renewcommand{\]}{\end{equation}}
\makeatother

\begin{document}
\title{Perturbative study of wave function evolution from source to detection
of a single particle and the measurement}
\author{Li Hua Yu}
\affiliation{Brookhaven National Laboratory}
\date{2/5/2025}
\begin{abstract}
We propose an experiment to perturb the evolution of the wave function
when a particle propagates through free space from a thin entrance
slit to a detector behind a thin exit slit. We consider inserting
a thin pin parallel to the two slits to perturb the system by blocking
the wave function, studying its perturbative effect on the detection
counting rate in a repeated experiment. Without the pin, the counting
rate is constant, and the 1D-Schr\textcyr{\"\cyro}dinger equation
gives the wave function evolution until the final detection. When
we insert a pin between the two slits, the change in the counting
rate is a function of the longitudinal and transverse positions $z$
and $x$ of the pin, defined as a perturbative function. We show that
the perturbative function is the functional derivative of the counting
rate; it is a real-valued function with phase information. The width
of the function starts from the entrance slit as the narrow width
of the entrance slit. It grows wider until it reaches a maximum and
then shrinks narrower and finally converges into the exit slit where
the particle is detected. This function has a spindle shape with its
pointed ends at the two slits. Hence it is very different from but
closely related to the well-known wave function of the Schr\textcyr{\"\cyro}dinger
equation with the initial condition at the entrance slit, which is
narrow only at the beginning, then grows wider until it reaches the
exit slit, where it is much larger than the slit width. We show that
the integral of the perturbative function over transverse position
x is a constant independent of z and equals twice the final counting
rate. This result provides information about when and where the wave
function collapses. While the image of the function has high resolution
with phase information, it requires only a low-intensity beam.
\end{abstract}
\email{lhyu@bnl.gov}

\maketitle

\section*{1. Indtroduction}

As pointed out by R. Penrose \cite{penrose}\cite{wiki_wavefunction_collapse},
\textquotedblleft The fundamental problem of quantum mechanics, as
that theory is presently understood, is to make sense of the reduction
of the state vector (i.e., the collapse of the wavefunction).'' \textquotedblleft This
issue is usually addressed in terms of the quantum measurement problem,
which is to comprehend how, upon measurement of a quantum system,
this (seemingly) discontinuous process can come about.'' According
to S. Weinberg \cite{weinberg}, in this problem, \textquotedblleft the
difficulty is not that quantum mechanics is probabilistic...The real
difficulty is that it is also deterministic, or more precisely, that
it combines a probabilistic interpretation with deterministic dynamics.\textquotedblright{}
\textquotedblleft This leaves the task of explaining them by applying
the deterministic equation for the evolution of the wavefunction,
the Schr�dinger equation, to observers and their apparatus.'' Bohr
offered an interpretation that is independent of a subjective observer,
or measurement, or collapse; instead, an \textquotedblleft irreversible''
or effectively irreversible process causes the decay of quantum coherence
which imparts the classical behavior of \textquotedblleft observation''
or \textquotedblleft measurement'' \cite{bohr,measurement_wiki}.
The important question is how are the probabilities converted into
an actual, well-defined classical outcome? In quantum mechanics, the
measurement problem is the problem of how, or whether, wavefunction
collapse occurs \cite{measurement_wiki} in this process. Gerard 't
Hooft in his discussion \cite{tHooft,phyiscs_today} about quantum
mechanics, pointed out that ``In practice, quantum mechanics merely
gives predictions with probabilities attached. This should be considered
as a normal and quite acceptable feature of predictions made by science:
different possible outcomes with different probabilities. In the world
that is familiar to us, we always have such a situation when we make
predictions. Thus the question remains: What is the reality described
by quantum theories? I claim that we can attribute the fact that our
predictions come with probability distributions to the fact that not
all relevant data for the predictions are known to us, in particular
important features of the initial state.'' 

R. Penrose \cite{penrose} summarized various points of view regarding
the fundamental problem of quantum mechanics, i.e., the wave function
collapse. The points of view in the first category we considered among
these \textquotedblleft do not normally take the view that any experimentally
testable deviations from standard quantum mechanics can arise within
these schemes.\textquotedblright{} In another category, \textquotedblleft against
all these are proposals of a different nature, according to which
it is argued that present-day quantum mechanics is a limiting case
of some more unified scheme,\textquotedblright{} such as that given
in \cite{pearle1,pearle2,Bialynicki,Ghirardi,Weinberg,Diosi}, as
referenced in \cite{penrose}, and the one proposed by R. Penrose
\cite{penrose}.

One question among these that attracts our particular interest most
is when and where the wave function collapses in a measurement experiment.
In particular, we would like to study whether, within the framework
of the bases of quantum mechanics, we can design an experiment to
find when, where, and in what sense the wave function collapses. We
limit our consideration to only the wave function collapse of a single
particle, so this experiment is as simple as possible. If the experiment
confirms the analysis, the precision of the measurement would be able
to provide an upper limit of possible deviation from the first category
of proposals and hence provide some experimental data for the second
category of proposals. For example, the error bar of the time scale
of the wave function collapse in the experiment would constrain the
parameters in the second category of theory.

``In quantum mechanics, wave function collapse, also called reduction
of the state vector \cite{penrose,wiki_wavefunction_collapse}, occurs
when a wave function\textemdash initially in a superposition of several
eigenstates\textemdash reduces to a single eigenstate due to interaction
with the external world.'' To study the wave function collapse in
an experiment for a particle to propagate from a source point to the
point where the detector is, we need to identify where the discontinuity
in the wave function is. We cannot directly measure the wave function
with a second detector between the source and the detector because
the detection will already cause a wave function collapse before it
reaches the detector. However, we can perturb the wave function in
the middle and measure the effect of the perturbation on the detection
rate. The detection rate is a functional of the wave function between
the source and the detector. Hence, we can use this fact to find information
about the middle of the propagation using functional derivative. The
perturbed function is not the wave function but is closely related
to it. And, as the perturbation approaches zero, the effect also approaches
zero. Hence, the functional derivative, which we defined as a perturbative
function, provides detailed information about the wave function as
if it were not perturbed.

Since the wave function must follow the Schr�dinger equation between
the source and the detector, the perturbative function must be continuous
between them. Discontinuity can only occur at either the source point
or the detector. Many consider the vector reduction to happen at the
detector during the measurement. However, when we examine this question
in detail, it is not evident that the discontinuity occurs at the
end. Our analysis of the experiment we proposed in this work shows
that discontinuity is at the beginning, i.e., at the source. For this,
we need to calculate the perturbative function. In the Heisenberg
picture, the wave function does not change; it remains the initial
wave function; hence, we analyze the evolution of the wave function
in the Schr�dinger picture.

Based on this understanding, this paper proposes an experiment to
provide information that will help address these questions \cite{yu1}.
In the next section, we will describe and outline our qualitative
analysis of the experiment. Then, following the next section, we will
present more detailed results introduced here quantitatively. Since
the derivation is straightforward and entirely based on the basics
of quantum mechanics, we leave it in the appendices.

\section*{2. Description of the Experiment, Qualitative Outline of the More
Detailed Analysis in the Next Sections}

As illustrated in Figure 1, we analyze the wave function evolution
of a particle when it propagates through free space in the longitudinal
$z$-direction from a thin entrance slit 1 to a detector behind a
thin exit slit 2 to find the information about whether the wave function
collapse occurs at the entrance slit 1 or the exit slit 2. The slits
are parallel to the $y$-axis (perpendicular to the plane of the figure).
The $x$-axis is vertical in the figure. Between the slits, the wave
function must follow the Schr�dinger equation because the only non-unitary
(irreversible) process is at the slits. Only the particles that pass
through the slits are selected and detected. The probability of a
particle found between $z_{1}$ and $z_{2}$ is a constant independent
of $z$ due to particle conservation. We insert a thin pin between
the two slits at position $z,x_{p}$ to cut off the wave function.
The caption for Fig. 1 gives the notations. 

As the pin width approaches zero, its effect on the counting rate
also approaches zero; their ratio gives the functional derivative
of the counting rate, i.e., the perturbative function as a function
of the $z,x_{p}$ that does not approach zero. Thus, the pin is a
perturbation to the wave function during the propagation. The result
provides information about the wave function evolution as if there
were no perturbation. In this section, we first qualitatively outline
the calculation of the perturbative function for the experiment in
Fig. 1, present the result, and discuss its physical meaning. A more
detailed quantitative calculation result will be in the following
sections.

\begin{figure*}
\includegraphics[width=0.6\columnwidth]{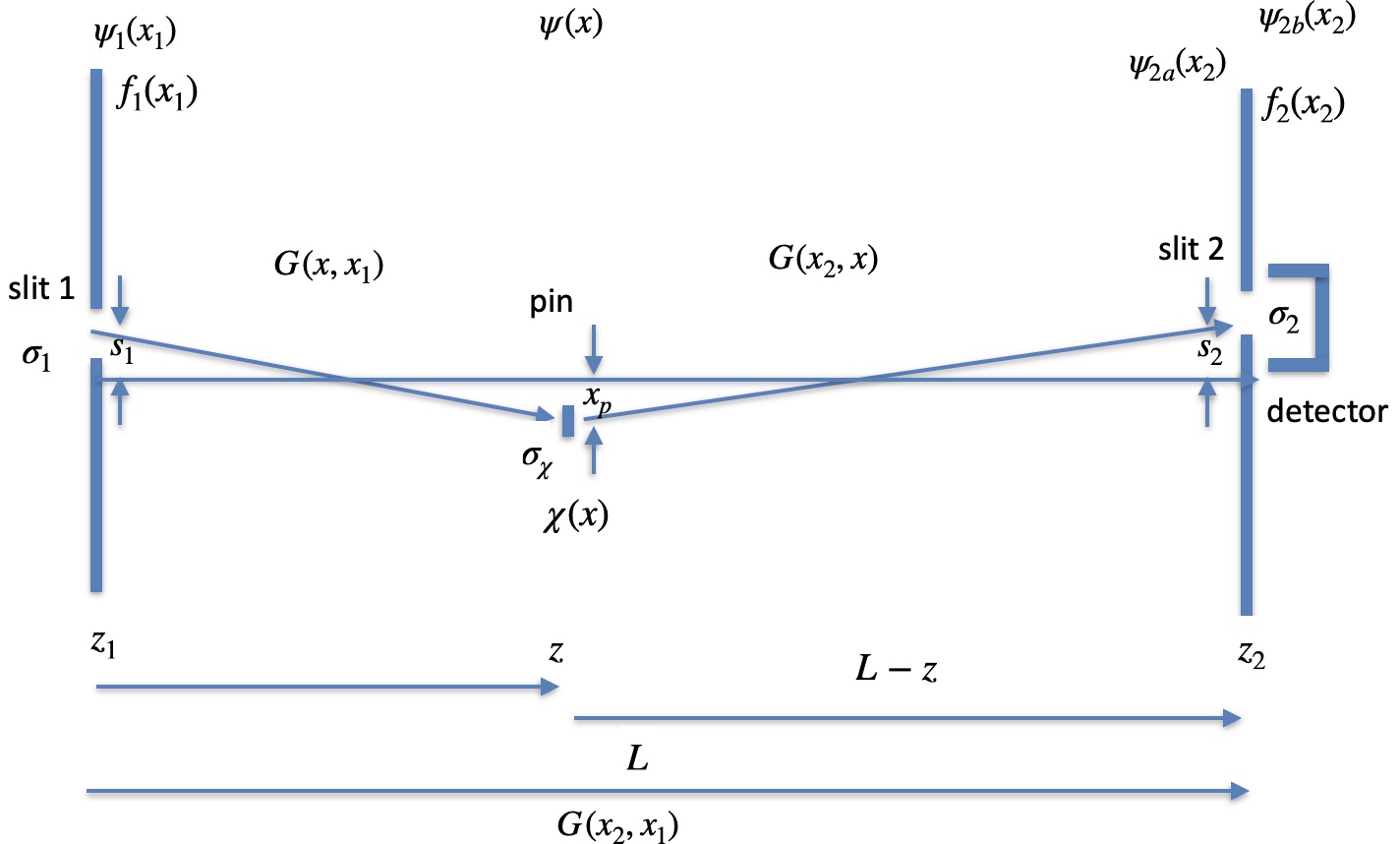}\vspace{-1.0em}

\caption{The wave functions at slits 1,2 at $z_{1},z_{2}$ with apertures $\sigma_{1},\sigma_{2}$
and corresponding transverse displacements $s_{1},s_{2}$ in x-direction,
and a pin at $z,x_{p}$ of width $\sigma_{\chi}$. The distances between
the slits and the pin are $z,L-z,L.$ The longitudinal is the $z$-axis,
the x-axis is vertical in this figure. The slits are perpendicular
to the plane of the figure, parallel to the $y$-axis. We use $\psi_{1}(x_{1},z_{1})$,
$\psi(x,z),$$\psi_{2a}(x_{2},z_{2})$ and $\psi_{2b}(x_{2},z_{2})$
to represent the wave function at the entrance, the pin and the exit,
respectively. The subscripts $a$ and $b$ represent before and after
slit 2. We use $f_{1}(x_{1})$ and $f_{2}(x_{2})$ to represent the
effect of the slits such that $\psi_{1}(x_{1},z_{1})\equiv f_{1}(x_{1}),\psi_{2b}(x_{2},z_{2})=f_{2}(x_{2})\psi_{2a}(x_{2},z_{2})$.
If we choose the slit with the hard-edged opening, $f_{1}(x_{1})$
and $f_{2}(x_{2})$ would be zero outside the slits and equal to 1
within the slits. To simplify the calculation, we assume they are
Gaussian with peak value 1, except that we choose $f_{1}(x_{1})$
to normalize $\psi_{1}$ as $P_{1}=\int dx_{1}|\psi_{1}(x_{1},z_{1}=0)|^{2}=1$.
The pin profile is $\chi(x)=1$ when it is removed. When inserted,
$\chi(x)=1-\exp(-\frac{1}{2\sigma_{\chi}^{2}}\left(x-x_{p}\right)^{2})$;
effective width (equivalent hard-edged slit width) is $\Delta x=\sqrt{2\pi}\sigma_{\chi}$.}
\vspace{-1.0em}
\end{figure*}

\subsection*{In section 3 we discuss 1D-Schr\textcyr{\"\cyro}dinger equation and
the propagation of wave function}

For simplicity, we neglect the effect of the particle spin, which
could be an electron or photon. When we use numerical examples, we
consider photons of $0.5\mu m$ wavelength. In section 3, we show
that when the slits are long and thin, under the paraxial approximation,
the wave function evolution follows the non-relativistic 1D-Schr�dinger
equation� until the final detection. The variables in the wave function
are $x$ and $t$. Because $z=vt$ where v is the particle velocity,
it is linear with $t$, we take $z$ as time. For a photon, $v=c$
is the light speed. The wave function is independent of $y$. We present
the well-known solution, the Green's function $G(x,x_{1})$, the wave
function $\psi(x),\psi_{2a}(x),\psi_{2b}(x)$ of the Schr�dinger equation
in Fig. 1, and the probability $P_{2b}$ of detection at the exit
when the entrance probability is normalized to one, which is proportional
to the counting rate. 

\subsection*{In Section 4 we derive Perturbative function $\frac{\delta}{\delta\chi(x,z)}P_{2b}$
and its integral $\frac{1}{2}\intop_{-\infty}^{\infty}dx\frac{\delta}{\delta\chi(x,z)}P_{2b}$
and discuss its physical meaning}

We study the perturbative effect of a pin on the counting rate $P_{2b}$
when $\Delta x$ is sufficiently small. This analysis generates a
function we denote as a perturbative function $\frac{\delta}{\delta\chi(x,z)}P_{2b}$.
This function provides information for the wave function evolution
process right before the detection. We find that $\frac{\delta}{\delta\chi(x,z)}P_{2b}$
is the functional derivative of detection probability $P_{2b}=\int dx_{3}|\psi_{2b}(x_{2},z_{2})|^{2}$
with respect to the perturbation $\Delta\chi$ of the pin at position
$(x,z)$ whithout the pin, i.e. $\chi(x)=1$. If at a point $x,$$\frac{\delta}{\delta\chi(x,z)}P_{2b}>0$,
the contribution of $\Delta x$ is $\Delta P_{2b}=\Delta x\frac{\delta}{\delta\chi(x,z)}P_{2b}$
gives the increase of the counting rate, then the pin decreases the
counting rate given by $P_{3b}$. If $\frac{\delta}{\delta\chi(x,z)}P_{2b}<0$,
the pin blocking at $x$ will increase the counting rate. We show
the function is real-valued, with both amplitude and phase information,
and is closely related to the wave function and proportional to the
product of two Feynman propagators \cite{Feynman} $G(x_{2},x)G(x,x_{1})$,
i.e., the probability amplitude of the path integral from $x_{1}$
passing through $x$ to reach $x_{2}$. Here, $G(x,x_{1})$ is the
Green's function of the Schr\textcyr{\"\cyro}dinger equation, the
transition probability amplitude from $x_{1},t_{1}$ to $x,t$, i.e.,
the Feynman propagator.

We compare $\frac{\delta}{\delta\chi(x,z)}P_{2b}$ (green) with $|\psi(x)|^{2}$
(red), and $\text{Re (}\psi(x))$ (magenta) for $z=1.95m$ in Fig.
2(a), where $s_{2}=3mm,z_{2}=2m$. The phase of $\frac{\delta}{\delta\chi(x,z)}P_{2b}$
advances with respect to $x$ very fast when $x<2mm$ or $x>4mm$.
Its advance rate becomes slow for $2mm<x<4mm$. It is zero at the
centroid $x=x_{c0}\approx2.92mm$ where we use a blue vertical line
to mark the point. Because of the fast phase advance, the green curve
has a very high oscillation frequency and is hardly visible except
near $x_{c0}$. Near $x_{c0}$, the phase advance is slow, the oscillation
frequency is low, and the fringes are relatively easy to detect in
the experiment. We also observe that the oscillation frequency for
$\frac{\delta}{\delta\chi(x,z)}P_{2b}$ in the $x$ direction is much
higher than that of $\text{Re(}\psi(x))$, and that we do not have
a way to measure $\psi(x)$ but $\frac{\delta}{\delta\chi(x,z)}P_{2b}$
can be measured.

\begin{figure*}
\includegraphics[width=0.6\columnwidth]{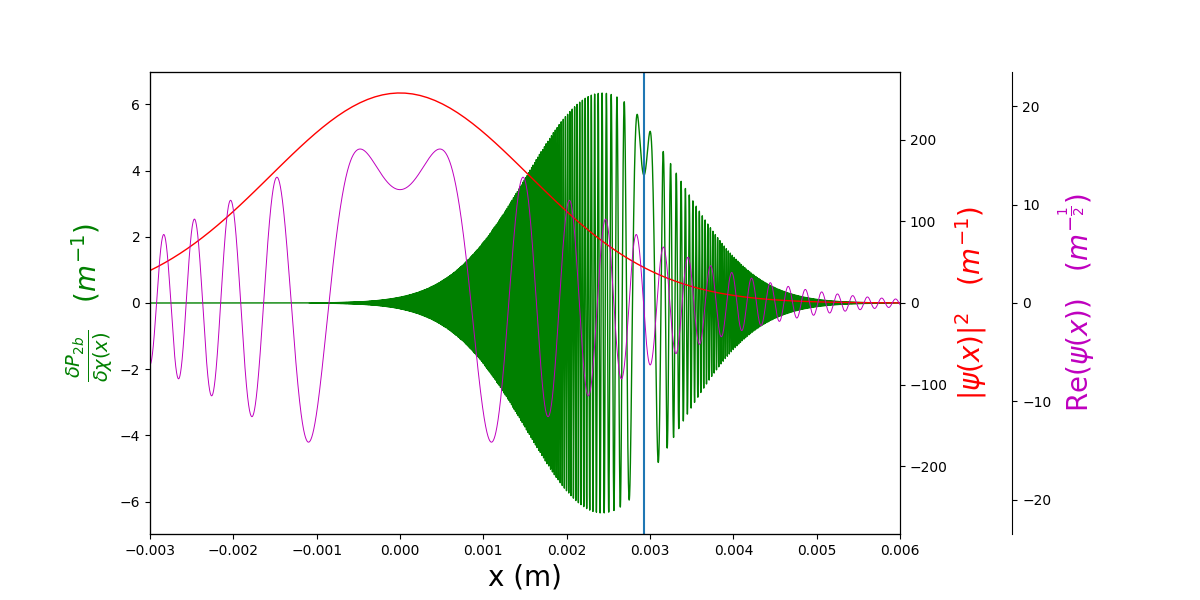}\vspace{-1.0em}\includegraphics[width=0.4\columnwidth]{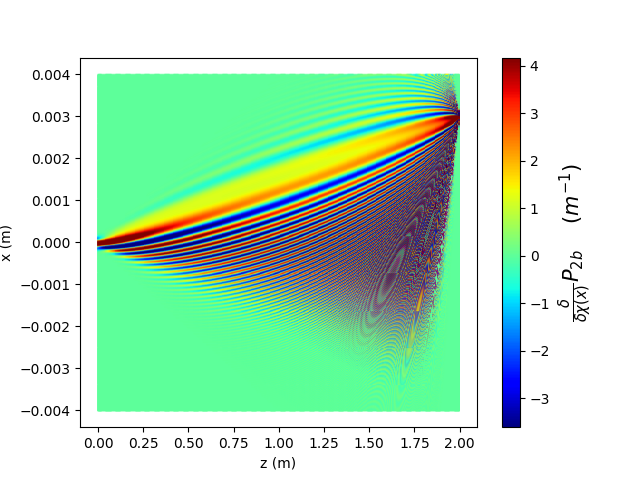}

\caption{(a) Compare the function $\frac{\delta P_{2b}}{\delta\chi(x,z)}$
with $|\psi(x,t)|^{2}$ (red), and $\text{Re (}\psi(x))$ (magenta)
, $s_{2}$=$3mm$, $z=1.95m$. (b) $\frac{\delta P_{2b}}{\delta\chi(x,z)}$
vs. $x,z$ in color scale}
\vspace{-1.0em}
\end{figure*}

Our calculation shows that $\frac{1}{2}\intop_{-\infty}^{\infty}dx\frac{\delta}{\delta\chi(x,z)}P_{2b}=P_{2b}$
is independent of $z$. \textbf{Thus, the physical meaning of the
perturbative function $\frac{1}{2}\frac{\delta P_{2b}}{\delta\chi(x,z)}$
is the detection probability contributed by the wave function per
unit section dx at point $(x,z)$. The contribution is linear and
measured in an experiment without using its absolute square.}

For the example in Fig. 2, $P_{2b}=0.0004258$. It is just half the
area between the green curve and the $x=0$ axis. The dominating contribution
to the area comes from the area near a centroid $x_{c0}$. Far from
$x_{c0}$, the positive and negative parts of the green curve cancel
each other, so their contribution to the probability is negligible.
We estimate the slow phase area is between $x_{c0}\pm\frac{1}{2}x_{\pi}$,
where $x_{\pi}$ is the distance of the first point away from $x_{c0}$
by a phase difference of $\pi$. In Fig. 3, we plot $x_{c0}(z)\pm\frac{1}{2}x_{\pi}(z)$
as a function of $z$ for slit 2 position $s_{2}=0,1,2,3$ mm, respectively.
The contour of the perturbative function $\frac{\delta}{\delta\chi(x,z)}P_{2b}$
for $s_{2}=3$ mm in Fig. 3 (the red contour) scans an area that corresponds
to the spindle-shaped bright area (colored yellow) in Fig. 2(b) with
its pointed ends at the two slits. In this area, the width $x_{\pi}$
is narrow from the beginning at the entrance slit. It grows wider
until it reaches a maximum and then shrinks narrower and finally collapses
into the exit slit where the particle is detected. Fig. 3 shows different
spindle-shaped areas, each for a different $s_{2}$. 

Hence the perturbative function $\frac{\delta}{\delta\chi(x,z)}P_{2b}$
is very different from the well-known wave function solution $\psi(x,z)$
of the Schr\textcyr{\"\cyro}dinger equation: with the initial condition
$\psi(x,z=z_{1})=\psi_{1}(x_{1}=x,z_{1})$ at the entrance slit 1,
$\psi(x,z)$ is only narrow at the beginning, then growing wider until
it reaches the exit slit where $\psi(x,z=z_{2})$ is much larger than
the width of exit slit 2 (see the blue curve in Fig. 3). In Fig. 3,
we compare the contours of the perturbative function with the RMS
width of $|\psi(x,z)|^{2}$. For this example, the photon wavelength
is $\lambda=0.5\mu m$, and $z_{2}=2m$, $\sigma_{1}=50\mu m$, $\sigma_{3}=4\mu m$.
From Fig. 3, we see that $x_{c0}(z)\pm\frac{1}{2}x_{\pi}(z)$ describes
specific events for particles detected at various exit slits at $s_{2}=0,1,2,3$
mm , while the well-known wave function $\psi(x,z)$ is the probability
amplitude; it describes the probability distribution along the entire
$x$-axis. The functions $x_{c0}(z)\pm\frac{1}{2}x_{\pi}(z)$ and
$\psi(x,z)$ have very different physical meanings and shapes.

Because of probability conservation, the total probability $P=\intop_{-\infty}^{\infty}dx|\psi(x,z)|^{2}=1$
is independent of $z$. Similarly, the probability of the particle
wave packets passing through each area is given by $\frac{1}{2}\intop_{-\infty}^{\infty}dx\frac{\delta}{\delta\chi(x,z)}P_{2b}(s_{2})=P_{2b}(s_{2})$.
It is also independent of $z$ when we normalize the entrance probability
to one.

\begin{figure*}
\includegraphics[width=0.4\columnwidth]{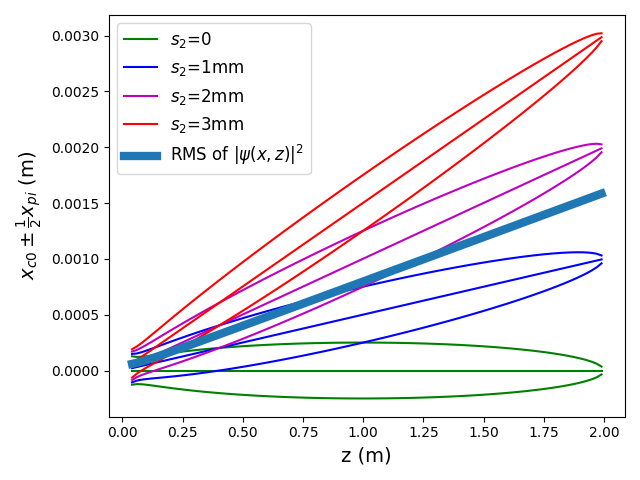}

\caption{The contours of $\frac{\delta P}{\delta\chi(x,z)}$, $x_{c0}\pm\frac{1}{2}x_{\pi}$
vs. $z$ for $s_{2}=0,1mm,2mm,3mm$}
\vspace{-1.0em}
\end{figure*}

Thus, once the experiment agrees with the prediction from $\frac{\delta}{\delta\chi(x,z)}P_{2b}$,
anywhere between slit 1 and 2, even for $z=0_{+}$ right after the
beginning, the total probability is already determined as $P_{2b}=\frac{1}{2}\intop_{-\infty}^{\infty}dx\frac{\delta}{\delta\chi(x,z)}P_{2b}$,
i.e., the final detection probability. Without the pin, the \textquotedblleft irreversible
process,\textquotedblright{} as a non-unitary process, can only occur
at either slit 1 or 2. For the example of $s_{2}=3$mm, $P_{1}=1$
at the entrance slit 1, and it immediately jumps to $P_{2b}=0.0004258$
at $z=0_{+}$ hence the probability collapses at slit 1. Thus, we
can divide the process into two parts: (1). The conceptual wave function
collapses, or, in other words, the probability distribution from total
probability 1 collapses to one specific individual event with the
probability $P_{2b}=0.0004258$, i.e., one specific case (i.e., the
detection by the detector at 3 mm) from a statistical distribution
is selected; (2). The continuous physical evolution process: from
the start, $\frac{\delta}{\delta\chi(x,z)}P_{2b}$ grows wider and
then becomes narrower until it finally collapses into the exit slit.
If the exit slit is sufficiently narrow, we take this as the wave
packet collapses into a point. This process appears closely related
to the discussion of t' Hooft {[}7{]} we mention at the beginning.

However, we can only test that for $z=0+\epsilon$, there is an agreement
with the prediction from $\frac{\delta}{\delta\chi(x,z)}P_{2b}$.
$\epsilon$ here is the distance from $z$ to slit 1. Fig. 2 shows
that close to slit1, the width of $\frac{\delta}{\delta\chi(x,z)}P_{2b}$
is tiny; the measurement error of $\frac{\delta}{\delta\chi(x,z)}P_{2b}$
becomes large, and hence there is a limit for $\epsilon$. Thus, if
there is a finite time scale for the wave function collapse for an
objective-collapse theory \cite{objective_collapse}, $\epsilon$
provided by an experiment may hint at the selection of some parameters
in the theory.

\subsection*{In Section 5 we describe wave packet propagation in an interference
experiment}

To demonstrate the phase information inherent in $\frac{\delta}{\delta\chi(x,z)}P_{2b}$,
we study an interference experiment as shown in Fig. 4. We replace
the single entrance slit 1 in Fig. 1 by two slits with $\psi_{1}(x_{1})=\frac{\left(\psi_{1+}(x_{1})+\psi_{1-}(x_{1})\right)}{\sqrt{2}}$,
where $\psi_{1+}(x_{1}),\psi_{1-}(x_{1})$ are Gaussian with the same
width $\sigma_{1}$ but centered at $s_{1}=1mm$ and $s_{1}=-1mm$
respectively, and they are in phase. Their separation $2s_{1}\gg\sigma_{1}$,
so their overlap is negligible, and hence the normalization $P_{1}=\int dx_{1}|\psi_{1}(x_{1},z_{1}=0)|^{2}=1$
is satisfied. The other parameters are the same as in Fig. 2. Fig.
4(a) shows $P_{2b}=\frac{1}{2}\intop_{-\infty}^{\infty}dx\frac{\delta}{\delta\chi(x,z)}P_{2b}$
calculated at $z=1.9m$ as the well-known interference parttern at
$z_{2}=L$. However, as discussed above, the plot is independent of
$z$: it is the same for any $0<z<2m$. In Fig. 4(b), when we choose
$s_{2}=0$, shown as the red dot in Fig. 4(a), where $P_{2b}=0.004122$
reaches the peak, there are two spindle-shaped regions symmetrically
oriented where $\frac{\delta}{\delta\chi(x,z)}P_{2b}>0$ varies slowly,
and both colored as between yellow to deep red, the two regions are
in phase. In Fig.4(c), we choose $s_{2}=-0.25mm$, shown as the green
dot in Fig. 4(b), where $P_{2b}=1.249\times10^{-5}$ reaches the minimum;
the two spindle-shaped regions have different colors. In the upper
plane, the region is colored blue. The lower region is colored red,
with phases differing by $\pi$.

\begin{figure*}
\includegraphics[viewport=14.40625bp 0bp 461bp 346bp,clip,scale=0.35]{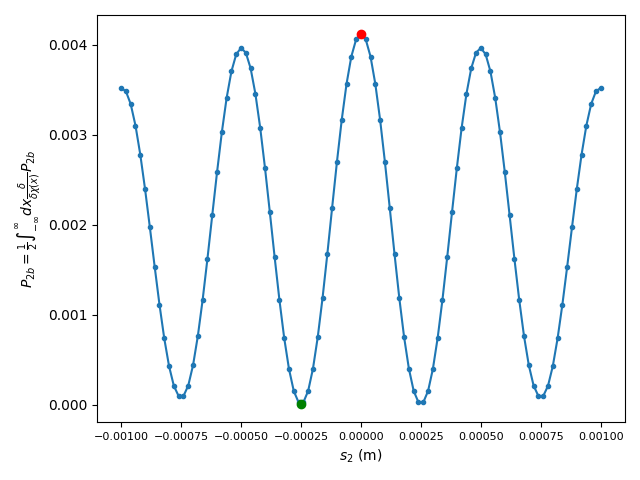}\includegraphics[viewport=0bp 0bp 439.391bp 346bp,clip,scale=0.4]{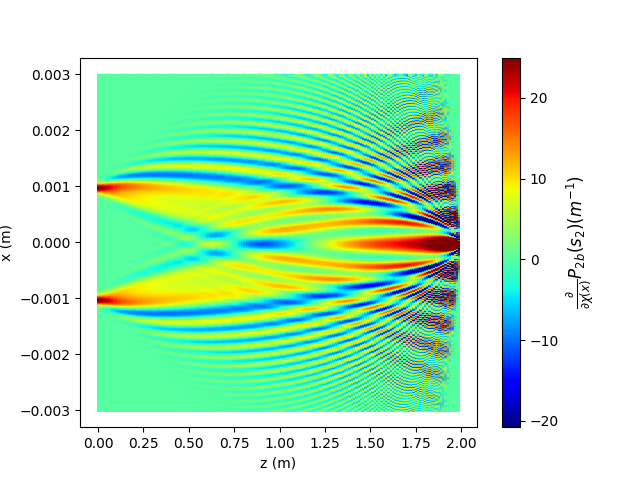}\includegraphics[scale=0.4]{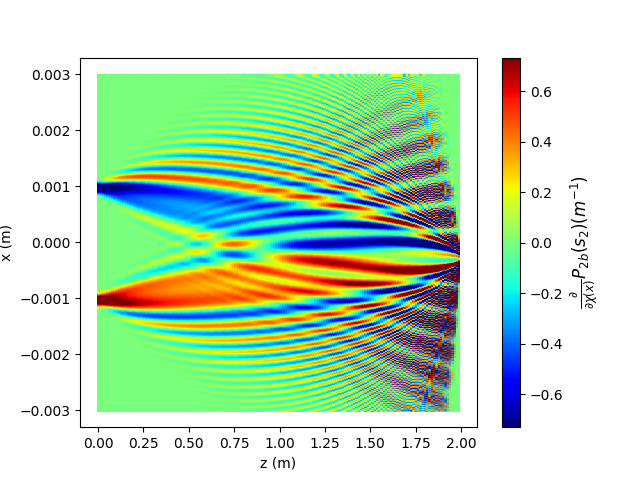}

\caption{(a) Interference pattern of $P_{2b}(s_{2})$ at $z_{2}=2m$ detector,
for $\lambda=0.5\mu m$ . (b).$\frac{\delta}{\delta\chi(x,z)}P_{2b}$
vs. $z,x$, at maximum $P_{2b}=0.004122$ at $s_{2}=0$. (c) $\frac{\delta}{\delta\chi(x,z)}P_{2b}$
at minimum $s_{2}=-0.25mm$ ${\displaystyle P_{2b}=1.249\times10^{-5}}$}
\vspace{-1.0em}
\end{figure*}

\subsection*{Outline of more quantitative description of the details of the calculation
and results discussed in this section}

Section 3 shows that the wave function evolution discussed in Fig.
1 follows the non-relativistic 1D-Schr\textcyr{\"\cyro}dinger equation
under the paraxial approximation until the final detection. We apply
the Green's function of the 1D-Schr\textcyr{\"\cyro}dinger equation
in steps to find the expressions of $\psi,\psi_{2a},\psi_{2b}$ from
the initial wave function $\psi_{1}$, and the detection probability
$P_{2b}$, which is proportional to the counting rate. The result
is presented in Sections 3.2 and 3.3, while the derivation is in Appendices
I and II.

In Section 4, we show that the perturbative probe pin scan generates
the functional derivative of $P_{2b}$, i.e. $\frac{\delta}{\delta\chi(x,z)}P_{2b}$.
Then we derive the analytical expression for $\frac{\delta}{\delta\chi(x,z)}P_{2b}$
by applying the expression of wave functions presented in Section
3. The result is used in Fig. 2. In the derivation, we show that $\frac{1}{2}\intop_{-\infty}^{\infty}dx\frac{\delta}{\delta\chi(x,z)}P_{2b}=P_{2b}$
is independent of $z$, and $\frac{\delta}{\delta\chi(x,z)}P_{2b}$
is proportional to the product of the Green's function $G(x_{2},x)G(x,x_{1})$
when the width $\sigma_{1}$ and $\sigma_{2}$ both approach zero,
i.e., the probability amplitude from $x_{1}$ passing through $x$
to $x_{2}$. We show that when $\sigma_{1},\sigma_{2}$ approach zero,
the limit of $P_{2b}$ has a simple expression. From the expression
of $\frac{\delta}{\delta\chi(x,z)}P_{2b}$ we derived its first fringe
width $x_{\pi}(z)$ and the centroid $x_{c0}(z)$, as shown in Fig.
3. One can use these expressions to check with the experiment. Then,
we estimate the required counting number to achieve a specified precision
of $\frac{\delta}{\delta\chi(x,z)}P_{2b}$ in the experiment. 

In Section 5, we briefly outline the calculation of the interference
patterns in Fig. 4 to show the importance of the phase information
provided by the function $\frac{\delta}{\delta\chi(x,z)}P_{2b}$ . 

In Section 6, we discuss our understanding regarding a few questions
about the wave function collapse raised and suggested by the analysis
above.

Section 7 is the summary.

\section*{3. 1D-Schr{\"O}dinger Equation And The Propagation Of
Wave Function}

\subsection*{3.1. Paraxial approximation of 2D-Schr\textcyr{\"\cyro}dinger equation
and Maxwell equation}

We first study the wave function evolution between the slits in Fig.
1. Between slits the wave function $\phi(x,z,t)$ must follow the
Schr\textcyr{\"\cyro}dinger equation,

\begin{align}
 & i\hbar\frac{\partial}{\partial t}\phi(x,z,t)=\left(\frac{P_{x}^{2}}{2M}+\frac{P_{z}^{2}}{2M}\right)\phi(x,z,t)=\left(-\frac{\hbar^{2}}{2M}\frac{\partial^{2}}{\partial x^{2}}-\frac{\hbar^{2}}{2M}\frac{\partial^{2}}{\partial z^{2}}\right)\phi(x,z,t)\label{2DshoedingerEq}
\end{align}

We do not include $y$ in the variables of the wave function because
we consider the system to be independent from $y$. $P_{x},P_{z}$
are the momentum operators.

Under the paraxial approximation (see e.g.,\cite{yariv,perdue,paraxial}),
we assume it can be approximated as a plane wave propagating in the
$z$-direction such that we can write it in the form $\phi(x,z,t)=\exp\left(i(kx-\omega_{k}t)\right)\psi(x,z)$,
where $\exp\left(i(kx-\omega_{k}t)\right)$ with particle energy $E_{k}=\hbar\omega_{k}=\frac{\hbar^{2}k^{2}}{2M}$
and momentum $p_{z}=\hbar k$ represents the plane wave, $kx-\omega_{k}t$
is its fast-varying phase, and $\psi(x,z)$ is its factor of slowly
varying amplitude and phase, so $\psi(x,z)$ represents the deviation
from the plane wave. Under the condition for this approximation $|k^{2}\psi|\gg|\frac{\partial^{2}\psi}{\partial z^{2}}|$,
$|2k\frac{\partial}{\partial z}\psi|\gg|\frac{\partial^{2}\psi}{\partial z^{2}}|$,
we have $P_{z}^{2}\phi=\hbar^{2}e^{i(kz-\omega_{k}t)}\left(k^{2}-2ik\frac{\partial}{\partial z}-\frac{\partial^{2}}{\partial z^{2}}\right)\psi$,
i.e. we can ignore the terms $-\frac{\partial^{2}\psi}{\partial z^{2}}$
compared with $k^{2}\psi-2ik\frac{\partial\psi}{\partial z}$ .

\begin{align}
 & \frac{P_{z}^{2}}{2M}\phi=\frac{\hbar^{2}}{2M}e^{i(kz-\omega_{k}t)}\left(k^{2}-2ik\frac{\partial}{\partial z}-\frac{\partial^{2}}{\partial z^{2}}\right)\psi\approx\frac{\hbar^{2}}{2M}e^{i(kz-\omega_{k}t)}\left(k^{2}-2ik\frac{\partial}{\partial z}\right)\psi\nonumber \\
 & i\hbar\frac{\partial}{\partial t}\phi(x,z,t)=e^{i(kz-\omega_{k}t)}\hbar\omega_{k}\psi=e^{i(kz-\omega_{k}t)}\frac{\hbar^{2}k^{2}}{2M}\psi\label{paraxial_approx}\\
 & \left(\frac{P_{x}^{2}}{2M}+\frac{P_{z}^{2}}{2M}\right)\phi(x,z,t)=e^{i(kz-\omega_{k}t)}\left(\frac{P_{x}^{2}}{2M}+\frac{\hbar^{2}k^{2}}{2M}-\frac{ik\hbar^{2}}{M}\frac{\partial}{\partial z}\right)\psi\nonumber 
\end{align}

Substituting into Eq.(\ref{2DshoedingerEq}), and using $v=\frac{\hbar k}{M}$
as the velocity of the particle with momentum $p_{z}=\hbar k=Mv$,
we get the 1D-Schr\textcyr{\"\cyro}dinger equation,

\begin{align}
 & i\hbar v\frac{\partial}{\partial z}\psi=-\frac{\hbar^{2}}{2M}\frac{\partial^{2}}{\partial z^{2}}\psi\label{1DschoedingerEq-1}
\end{align}

When we consider $z$ taking the role of time with $z=vt$, we get
the more familiar form of the 1D-Schr\textcyr{\"\cyro}dinger equation,

\begin{align}
 & i\hbar\frac{\partial}{\partial t}\psi=-\frac{\hbar^{2}}{2M}\frac{\partial^{2}}{\partial x^{2}}\psi\label{1DschoedingerEq}
\end{align}

We may rewrite the Eq.(\ref{1DschoedingerEq-1}) as $i\frac{\partial}{\partial z}\psi=\left(-\frac{\hbar}{2Mv}\frac{\partial^{2}}{\partial x^{2}}\right)\psi=-\frac{1}{2k}\frac{\partial^{2}}{\partial x^{2}}$,
then the only parameter in the experiment we discussed in Fig. 1 is
$k=\frac{2\pi}{\lambda}$. To calculate the functions listed in Fig.
1: $\psi_{2a},\psi_{2b}$ we only need to specify the de Broglie wavelength
$\lambda=0.5\mu m$ in addition to the specification of the slits
($z_{j}\equiv vt_{j},s_{j},\sigma_{j};j=1,2)$ to specify the experiment,
all in units of length. So the result can be expressed by a combination
of lengths and some dimensionless parameters of the ratio of lengths.
This helps to guide us to find simpler expressions.

Similarly, we assume a 2D Maxwell equation (because we do not consider
polarization, we just use $\phi$ to denote any component of the electromagnetic
field),

\begin{align}
 & \frac{1}{c^{2}}\frac{\partial^{2}}{\partial t^{2}}\phi(x,z,t)=\left(\frac{\partial^{2}}{\partial x^{2}}+\frac{\partial^{2}}{\partial z^{2}}\right)\phi(x,z,t)\label{maxwelleq}
\end{align}

Following the same way \cite{perdue}, we assume $k=\frac{\omega}{c}$,
$\phi(x,z,t)=e^{i(kz-\omega t)}\psi(x,z)$, take the paraxial approximation:
\textbar$\frac{\partial^{2}}{\partial z^{2}}\psi|\ll|\frac{\partial^{2}}{\partial x^{2}}\psi$\textbar ,
\textbar$\frac{\partial^{2}}{\partial z^{2}}\psi|\ll2k|\frac{\partial}{\partial z}\psi$\textbar ,
we get the same equation $i\frac{\partial}{\partial z}\psi=-\frac{1}{2k}\frac{\partial^{2}}{\partial x^{2}}\psi$.
Then, let $v=c,z=vt,p_{z}=k\hbar=Mv$, it is the same as Eq.(\ref{1DschoedingerEq})
again. Again, we still can characterize the experiment by specifying
$\lambda=\frac{2\pi}{k}$ only. 

\subsection*{3.2 Green's function and the propagation of wave function }

The Green's function of free space 1D-Schr\textcyr{\"\cyro}dinger
equation Eq.(\ref{1DschoedingerEq}) is in basic quantum physics \cite{schiff},

\begin{equation}
G(x_{2},x_{1};t_{2}-t_{1})=\left(\frac{M}{2\pi i\hbar(t_{2}-t_{1})}\right)^{\frac{1}{2}}\exp\left[i\frac{M}{2\hbar(t_{2}-t_{1})}\left(x_{2}-x_{1}\right)^{2}\right]\label{greens_function-1}
\end{equation}

We write another derivation of the Green's function in the Heisenberg
picture \cite{yu} in Appendix I because later in the Section 6, we
shall use the derivation process to explain why the position of the
detector at the exit slit can affect at the entrance slit the initial
wave packet that will evolve into the exit slit at the end.

The initial wave function at slit 1 is $\psi_{1}(x_{1})=f_{1}(x_{1})=\left(\frac{1}{2\pi\sigma_{1}^{2}}\right)^{\frac{1}{4}}\exp\left(-\frac{1}{4\sigma_{1}^{2}}(x_{1}-s_{1})^{2}\right)$,
normalized with $P_{1}=\int dx_{1}|\psi_{1}(x_{1},z_{1})|^{2}=1$.
For the wave function propagation from slit 1 to 2, we first calculate
the wave function $\psi_{2a}(x_{2},t_{2})$. After passing through
the slit 2, we have 

\begin{align}
 & \psi_{2a}(x_{2},z_{2})=\int dx_{1}G(x_{2},x_{1};t_{2}-t_{1})\psi_{1}(x_{1},z_{1})\label{propagation}\\
 & \psi_{2b}(x_{2},z_{2})=f_{2}(x_{2})\psi_{2a}(x_{2},z_{2})\nonumber 
\end{align}

where $f_{2}(x_{2})=\exp(\alpha_{f2}(q_{2}-s_{2})^{2}),\alpha_{f2}\equiv-\frac{1}{4\sigma_{2}^{2}}$.

Because $\ln G(x_{2},x_{1};t_{2}-t_{1}),\ln\psi_{1}(x_{1},z_{1})$
and $\ln f_{2}(x_{2})$ are quadratic forms, $\psi_{2a},\psi_{2b}$
are derived using Gaussian integral in Appendix II. The derivation
is in the basics of quantum mechanics, e.g., in \cite{schiff}. We
introduce parameters $\alpha_{1},\alpha_{2a},\alpha_{2b},c_{1},c_{2a},c_{2b},s_{2b},d,\Delta,d_{1},\Delta_{1}$
defined in terms of the experiment parameters $\sigma_{1},\sigma_{2},s_{1},s_{2},x_{1},z_{1},x,z,x_{2},z_{2}$
(see Fig. 1 for the notation) in the following to simplify the expression
for $\psi_{2a},\psi_{2b}$: 

\begin{align}
 & \frac{1}{\alpha_{1}}=-4\sigma_{1}^{2},d_{1}=\frac{\lambda z_{2}}{4\pi}=\frac{\lambda L}{4\pi},\Delta_{1}\equiv\sigma_{1}^{2}+id_{1}\nonumber \\
 & \frac{1}{\alpha_{2a}}=-4\Delta_{1},\frac{1}{\alpha_{2b}}=-4\sigma_{2}^{2}\frac{\Delta_{1}}{\Delta_{1}+\sigma_{2}^{2}},s_{2b}=\left(s_{2}-s_{1}\right)\frac{\Delta_{1}}{\Delta_{1}+\sigma_{2}^{2}}+s_{1}\label{parameters}\\
 & d=\frac{\lambda z}{4\pi},\Delta\equiv\sigma_{1}^{2}+id\nonumber 
\end{align}

The result is expressed in the following quadratic forms:

\begin{align}
 & \ln\psi_{1}(x_{1},z_{1})=\alpha_{1}(x_{1}-s_{1})^{2}+c_{1}=-\frac{1}{4\sigma_{1}^{2}}(x_{1}-s_{1})^{2}+\frac{1}{4}\ln\left(\frac{1}{2\pi\sigma_{1}^{2}}\right)\nonumber \\
 & \ln\psi_{2a}(x_{2},z_{2})=\alpha_{2a}\left(x_{2}-s_{1}\right)^{2}+c_{2a}=-\frac{1}{4\Delta_{1}}\left(x_{2}-s_{1}\right)^{2}+\frac{1}{2}\ln\left(\frac{\sigma_{1}^{2}}{\Delta_{1}}\right)+\frac{1}{4}\ln\left(\frac{1}{2\pi\sigma_{1}^{2}}\right)\nonumber \\
 & \ln\psi(x,z)=\ln\psi_{2a}(x_{2}=x,z_{2}=z)=-\frac{1}{4\Delta}\left(x-s_{1}\right)^{2}+\frac{1}{2}\ln\left(\frac{\sigma_{1}^{2}}{\Delta}\right)+\frac{1}{4}\ln\left(\frac{1}{2\pi\sigma_{1}^{2}}\right)\nonumber \\
 & \ln\psi_{2b}(x_{2},z_{2})=\alpha_{2b}\left(x_{2}-s_{2b}\right)^{2}+c_{2b}=-\frac{1}{4\Delta_{1}}\frac{\Delta_{1}+\sigma_{2}^{2}}{\sigma_{2}^{2}}\left(x_{2}-s_{2b}\right)^{2}-\frac{(s_{1}-s_{2})^{2}}{4\left(\Delta_{1}+\sigma_{2}^{2}\right)}+\frac{1}{2}\ln\left(\frac{\sigma_{1}^{2}}{\Delta_{1}}\right)+\frac{1}{4}\ln\left(\frac{1}{2\pi\sigma_{1}^{2}}\right)\label{psi_pattern}
\end{align}

where we used $\psi(x,z)=\psi_{2a}(x_{2}=x,z_{2}=z)$ which is obvious
when we examine the setup in Fig. 1.

To check this result, when using a specific example so that all the
coefficients of the quadratic forms in the intermediate steps are
numerical, the numerical check of the result using Eq.(\ref{propagation})
by Gaussian integration is straightforward and simple.

\subsection*{3.3 Detection probability $P_{2b}$ and probability density $|\psi(x,z)|^{2}$}

The expression of $\ln\psi_{2b}(x_{2},z_{2})$ in Eq.(\ref{psi_pattern})
is again a quadratic form of $x_{2}$, hence $\ln\psi_{2b}(x_{2},z_{2})+\ln\psi_{2b}^{*}(x_{2},z_{2})$
in the integrand of $P_{2b}=\int dq_{2}|\psi_{2b}(x_{2})|^{2}$ is
also a quadratic form. Thus, in Appendix II Eq.(\ref{gaussian_integral})
can be applied as a Gaussian integral; we derive $P_{2b}$ expressed
in terms of the scaled dimensionless parameters $\mu,\rho$ defined
in the following,

\begin{align}
 & P_{2b}=\sqrt{\frac{\rho\mu^{2}}{{\displaystyle \mu^{2}+\rho\mu^{2}+1}}}\exp\left(-\frac{1}{2\sigma_{1}^{2}}\frac{\mu^{2}}{\mu^{2}+\rho\mu^{2}+1}(s_{1}-s_{2})^{2}\right)\label{P2b}\\
 & \mu\equiv\frac{\sigma_{1}^{2}}{d_{1}}=\frac{4\pi\sigma_{1}^{2}}{\lambda L},\rho\equiv\frac{\sigma_{2}^{2}}{\sigma_{1}^{2}}\nonumber 
\end{align}

In the limit of small slit sizes, as $\sigma_{1},\sigma_{2}\rightarrow0$,
we need to specify their ratio $\rho=\frac{\sigma_{2}^{2}}{\sigma_{1}^{2}}$,
we find 

\begin{equation}
P_{2b}=\sqrt{{\displaystyle \rho}\mu^{2}}\exp\left(-\frac{\mu^{2}}{2\sigma_{1}^{2}}(s_{1}-s_{2})^{2}\right)=\frac{\sigma_{1}\sigma_{2}}{d_{1}}\exp\left(-\frac{\sigma_{1}^{2}}{2d_{1}^{2}}(s_{1}-s_{2})^{2}\right)\label{P2blimit}
\end{equation}

For the example in Fig. 2, $\sigma_{1}=50\mu m,\sigma_{2}=4\mu m,$
$s_{1}=0,s_{2}=3mm,$ for $z_{2}=L=2m$, $d_{1}=\frac{\lambda L}{4\pi}=\frac{0.5\mu m\times2m}{4\pi}=7.96\times10^{-8}m^{2}$,
and $\mu^{2}=\left(\frac{\sigma_{1}^{2}}{d_{1}}\right)^{2}=\left(\frac{2500\times10^{-12}}{7.96\times10^{-8}}\right)^{2}=9.9\times10^{-4}\ll1$,
the Eq.(\ref{P2blimit}) for small beam size limit gives a good estimation
near the detector. Compare the result of Eq.(\ref{P2b}) with Eq.(\ref{P2blimit})

\begin{align}
 & P_{2b}={\displaystyle 0.00042584}\label{P3bo}\\
 & P_{2b}\approx\frac{\sigma_{1}\sigma_{2}}{d_{1}}\exp\left(-\frac{\sigma_{1}^{2}}{2d_{1}^{2}}\left(s_{2}-s_{1}\right)^{2}\right)={\displaystyle {\displaystyle 0.00251}}\times{\displaystyle 0.16922}={\displaystyle 0.0004253}\nonumber 
\end{align}

From the experimental setup in Fig. 1, we see that $\psi(x,z)=\psi_{2a}(x,z)$
represents free space wave function propagation emitted from slit
1. From the probability amplitude $\psi(x,z)$ in Eq.(\ref{psi_pattern}),
we find 

\begin{equation}
|\psi(x,z)|^{2}={\displaystyle \frac{\sqrt{2}e^{-\frac{\sigma_{1}^{2}\left(x-s_{1}\right)^{2}}{2\left(d^{2}+\sigma_{1}^{4}\right)}}\left|\frac{\sigma_{1}^{2}}{\Delta}\right|}{2\sqrt{\pi}\sqrt{\left|\sigma_{1}^{2}\right|}}}\label{abs_psi2a}
\end{equation}

Hence the RMS width of $|\psi(x,z)|^{2}$ is $\sigma={\displaystyle \sqrt{\frac{d^{2}+\sigma_{1}^{4}}{\sigma_{1}^{2}}}}$
where $d=\frac{\lambda z}{4\pi}$, it is plotted in Fig. 3. $\text{\ensuremath{\sigma(z)}}$
increases with $z$ linearly when $d\gg\sigma_{1}^{2}$.

\section*{4. Perturbative function $\frac{\delta}{\delta\chi(x,z)}P_{2b}$
, its integral $\frac{1}{2}\intop_{-\infty}^{\infty}dx\frac{\delta}{\delta\chi(x,z)}P_{2b}$
and its physical meaning}

\subsection*{4.1 Calculate the contribution of a point between the source and
detection point to the final detection }

We calculate the contribution to $P_{2b}$ of a pin of infinitesimal
width $\delta x$ at $(x,z)$ between source point 1 and detection
point 2 according to the set up in Fig. 1. Applying the formulas in
Section 3, we have 

\begin{align}
 & P_{2b}=\int_{-\infty}^{\infty}dx_{2}|\psi_{2b}(x_{2})|^{2}\\
 & \psi_{2b}(x_{2})=f_{2}(x_{2})\int_{-\infty}^{\infty}dxG(x_{2},x;t_{2}-t)\chi(x)\int_{-\infty}^{\infty}dx_{1}G(x,x_{1};t-t_{1})f_{1}(x_{1})\nonumber 
\end{align}

Thus, we consider $P_{2b}$ a functional of the function $\chi(x)$
(the profile of the pin in Fig. 1) when we use a pin of width $\delta x$
to block the wave function at position $(x,z)$. Let $\chi(x)=1$.
The contribution of the section $\delta x$ to $P_{2b}$ is given
by the functional derivative of $P_{2b}$ as $\frac{\delta P_{2b}}{\delta\chi(x,z)}\delta x$,
hence the effect of the pin is $\delta P_{2b}=-\frac{\delta P_{2b}}{\delta\chi(x,z)}\delta x$.
Thus, we can measure $\frac{\delta P_{2b}}{\delta\chi(x,z)}$ using
the pin blocking. We have

\begin{align}
 & \frac{\delta P_{2b}}{\delta\chi(x,z)}=\frac{\delta}{\delta\chi(x,z)}\int dx_{2}|\psi_{2b}(x_{2})|^{2}=\int_{-\infty}^{\infty}dx_{2}\frac{\delta}{\delta\chi(x,z)}\left(\psi_{2b}(x_{2})\psi_{2b}^{*}(x_{2})\right)\label{pertubative_function}\\
 & =\int_{-\infty}^{\infty}dx_{2}\left(f_{2}(x_{2})G(x_{2},x;t_{2}-t)\int_{-\infty}^{\infty}dx_{1}G(x,x_{1};t-t_{1})f_{1}(x_{1})\right)\psi_{2b}^{*}(x_{2},z_{2})+c.c.\nonumber \\
 & =\left(\int_{-\infty}^{\infty}dx_{2}\psi_{2b}^{*}(x_{2},z_{2})f_{2}(x_{2})G(x_{2},x;t_{2}-t)\right)\int_{-\infty}^{\infty}dx_{1}G(x,x_{1};t-t_{1})\psi_{1}(x_{1})+c.c.\nonumber 
\end{align}

Since $\chi(x)=1$, $\chi(x)$ is already removed by the differentiation,
the experiment is equivalent to an experiment whithout the pin. Thus
we have

\begin{align}
 & \frac{\delta P_{2b}}{\delta\chi(x,z)}=\phi_{b}(x)\psi(x)+c.c.\label{perturbative_function}\\
 & \psi(x)\equiv\int_{-\infty}^{\infty}dx_{1}G(x,x_{1};t)\psi_{1}(x_{1})\nonumber \\
 & \phi_{b}(x)\equiv\int_{-\infty}^{\infty}dx_{2}\psi_{2b}^{*}(x_{2},z_{2})f_{2}(x_{2})G(x_{2},x;t_{2}-t)\nonumber 
\end{align}

Here we define another function $\phi_{b}(x_{2})$, it represents
the propagation from point $x$ to $x_{2}$. We let $z_{2}=L$, and
we have taken $t-t_{1}\equiv t=z/v$ and $t_{2}-t=(L-z)/v$, and the
pin position is $(x,z)$.

From this expression, we see that $\frac{\delta P_{2b}}{\delta\chi(x,z)}$
is closely related to Green's function $G(x_{2},x;t_{2}-t)G(x,x_{1};t-t_{1})$;
it is the product of the two Feynman propagators and represents the
path integral from $x_{1}$ to $x_{2}$ passing through point $x$,
or in other words, the corresponding probability amplitude. When $\sigma_{1},\sigma_{2}$
is finite (non-zero), $\frac{\delta P_{2b}}{\delta\chi(x,z)}$ is
a linear combination of $G(x_{2},x;t_{2}-t)G(x,x_{1};t-t_{1})$, a
sum over contribution from $x_{1},x_{2}$ of slit 1 and 2. It is a
Gaussian integral over $x_{1},x_{2}$.

\subsection*{4.2 The relation between $\frac{\delta P_{2b}}{\delta\chi(x,z)}$
and $P_{2b}$, $P_{2b}=\frac{1}{2}\int_{-\infty}^{\infty}dx\frac{\delta P_{2b}}{\delta\chi(x,z)}$
, and its relation to wave function collapse}

From Eq.(\ref{pertubative_function}) we have

\begin{align}
 & \int_{-\infty}^{\infty}dx\frac{\delta P_{2b}}{\delta\chi(x,z)}=\int_{-\infty}^{\infty}dx\int_{-\infty}^{\infty}dx_{2}\psi_{2b}^{*}(x_{2},z_{2})f_{2}(x_{2})G(x_{2},x;t_{2}-t)\int_{-\infty}^{\infty}dx_{1}G(x,x_{1};t)\psi_{1}(x_{1})+c.c.\nonumber \\
 & =\int_{-\infty}^{\infty}dx_{2}\psi_{2b}^{*}(x_{2},z_{2})f_{2}(x_{2})\int_{-\infty}^{\infty}dx_{1}\int_{-\infty}^{\infty}dxG(x_{2},x;t_{2}-t)G(x,x_{1};t-t_{1})\psi_{1}(x_{1})+c.c.\label{P3bo2}\\
 & =\int_{-\infty}^{\infty}dx_{2}\psi_{2b}^{*}(x_{2},z_{2})f_{2}(x_{2})\int_{-\infty}^{\infty}dx_{1}G(x_{2},x_{1};t_{2}-t_{1})\psi_{1}(x_{1})+c.c.\nonumber \\
 & =\int_{-\infty}^{\infty}dx_{2}\psi_{2b}^{*}(x_{2},z_{2})\psi_{2b}(x_{2},z_{2})+c.c.=\int_{-\infty}^{\infty}dx_{2}|\psi_{2b}(x_{2})|^{2}+c.c.=P_{2}+c.c.=2P_{2b}\nonumber 
\end{align}

Thus we have a simple relation between $\frac{\delta}{\delta\chi(x,z)}P_{2b}$
and $P_{2b}$, 

\begin{equation}
P_{2b}=\frac{1}{2}\int_{-\infty}^{\infty}dx\frac{\delta P_{2b}}{\delta\chi(x,z)}\label{PdPrelation}
\end{equation}

Thus, $\int_{-\infty}^{\infty}dx\frac{\delta P_{2b}}{\delta\chi(x,z)}$
is independent of $z=vt$, it is a constant, i.e., twice the probability
of a particle passing through the exit slit 2. Notice that we denote
the pin position as $(x,z)$. $P_{2b}=\frac{1}{2}\int_{-\infty}^{\infty}dx\frac{\delta P_{2b}}{\delta\chi(x,z)}$
is independent from $z$, meaning immediately after passing through
the entrance slit, the integral of $\frac{1}{2}\int_{-\infty}^{\infty}dx\frac{\delta P_{2b}}{\delta\chi(x)}$
has already been determined as $P_{2b}$ at the exit. This result
confirms the discussion in the Introduction about the process of wave
function collapse. 

\textbf{The perturbative function $\frac{1}{2}\frac{\delta P_{2b}}{\delta\chi(x,z)}dx$
has a straightforward physical meaning. It contributes to the detection
probability from a section dx at point $(x,z)$. The contribution
is linear, without the need to calculate its absolute square, and
can be measured in an experiment.}

We consider the slit function as Gaussian because the derivation of
the analytical expression is simple for a single slit. For multiple
detectors on the exit screen, the function for different slits will
have an overlap that is not negligible unless their separation is
much larger than $\sigma_{2}$. We can approximate the Gaussian slit
function $f_{2}(x_{2})=\exp(-\frac{1}{2\sigma_{2}^{2}}(x_{2}-s_{2})^{2})$
with a hard-edged slit of width $\Delta x_{2}=\sqrt{2\pi}\sigma_{2}$
when the slit width is small $\sqrt{2\pi}\sigma_{2}\ll\sigma_{2a},$
where $\sigma_{2a}={\displaystyle \sqrt{\frac{d_{1}^{2}+\sigma_{1}^{4}}{\sigma_{1}^{2}}}}$
in Eq.(\ref{abs_psi2a}) is the RMS width of $|\psi_{2a}(s_{2})|^{2}$
:

\begin{equation}
P_{2b}(s_{2})=\int_{-\infty}^{\infty}dx_{2}\exp(-\frac{1}{2\sigma_{2}^{2}}(x_{2}-s_{2})^{2})|\psi_{2a}(x_{2})|^{2}\approx|\psi_{2a}(s_{2})|^{2}\sqrt{2\pi}\sigma_{2},\label{harg_edge_approximation-1}
\end{equation}

For a hypothetical experiment, we may consider an infinite set of
slits $\{s_{2n}\}$, each with width $\Delta s_{2n}=\sqrt{2\pi}\sigma_{2}$
extending over the exit screen. In the limit $\sigma_{2}\rightarrow0$,
the total probability over the sum of all these slits is $\sum_{n}P_{2b}(s_{2n})$,
so it is

\begin{align}
 & \lim_{\sigma_{2}\rightarrow0}\sum_{n}\frac{\Delta s_{2n}}{\sqrt{2\pi}\sigma_{2}}P_{2b}(s_{2n})=\int_{-\infty}^{\infty}dx_{2}|\psi_{2a}(x_{2})|^{2}\int_{-\infty}^{\infty}\frac{ds_{2n}}{\sqrt{2\pi}\sigma_{2}}\exp(-\frac{1}{2\sigma_{2}^{2}}(x_{2}-s_{2n})^{2})=\int dx_{2}|\psi_{2a}(x_{2})|^{2}=1.\label{sn_probability}
\end{align}

For each $s_{2n}$, $P_{2b}(s_{2n})=\frac{1}{2}\int_{-\infty}^{\infty}dx\frac{\delta P_{2b}(s_{2n})}{\delta\chi(x)}$
gives the probability of the particle passing through the slit $s_{2n}$.
Thus, $\psi_{2a}(x_{2})$ represents the probability amplitude of
the particle distribution. As a comparison, $\frac{\delta P_{2b}(s_{2n})}{\delta\chi(x)}$
represents an individual single particle that eventually passes through
the slit at $s_{2n}$. The fact that $P_{2b}(s_{2n})=\frac{1}{2}\int_{-\infty}^{\infty}dx\frac{\delta P_{2b}(s_{2n})}{\delta\chi(x)}$
is independent of $z$ even though $\frac{\delta P_{2b}(s_{2n})}{\delta\chi(x,z)}$
does depend on $z$, means that even at the beginning at $z=0_{+}$,
the specific sampling from the distribution represented by $\psi_{2a}(x_{2})$
has already realized.

We shall use the independence from $z$ of Eq.(\ref{PdPrelation})
to simplify the derivation of $\frac{\delta}{\delta\chi(x,z)}P_{2b}=\frac{\delta P_{2bo}}{\delta\chi(x,z)}=\phi_{b}(x)\psi(x)+c.c.$
in Appendix IV, the result is given in section 4.3.

\subsection*{4.3 The perturbative function $\frac{\delta}{\delta\chi(x,z)}P_{2b}$
and the pin blocking probability change $\frac{\Delta P_{2b}}{P_{2b}}$
in the experiment}

Using Eq.(\ref{perturbative_function}) and Eq.(\ref{PdPrelation}),
we obtained the analytical expression for the perturbative function,
as shown in Appendix IV, Eq.(\ref{dPdchi-1}).

\begin{align}
 & \frac{\delta P_{2b}}{\delta\chi(x,z)}=P_{2b}\frac{\sqrt{-\alpha_{\chi}}}{\sqrt{\pi}}\exp(\alpha_{\chi}\left(x-x_{c}\right)^{2})+c.c.\label{dPdchi}
\end{align}

Where $\alpha_{\chi}$ and $x_{c}$ are derived and simplified in
Appendix IV, and $P_{2b}$ is in Eq.(\ref{P2b}). To make writing
simpler, defining the dimensionless parameter of distance ratio $\xi$
as follows, we find 

\begin{align}
 & \alpha_{\chi}=-\frac{1}{2\sigma_{1}^{2}}{\displaystyle \frac{i\mu{\displaystyle \left(\mu^{2}+\rho\mu^{2}+1\right)}}{\left(-i\mu+\xi\right)\left(\mu{\displaystyle \left(i\xi-\mu\right)}\rho+2\left(\xi-1\right)\left(i\mu+1\right)\right)}}\label{alphachi-2}\\
 & \xi\equiv\frac{d}{d_{1}}=\frac{z}{L}\nonumber 
\end{align}

where $\mu,\rho$ are defined in Eq.(\ref{P2b}), $d,d_{1}$ are defined
in Eq.(\ref{parameters}). $x_{c}$ is complex-valued, 

\begin{align}
 & x_{c}=\frac{{\displaystyle c_{s1}}s_{1}+{\displaystyle c_{s2}}s_{2}}{{\displaystyle \mu^{2}+\rho\mu^{2}+1}}\label{xc-2}\\
 & {\displaystyle c_{s1}}\equiv{\displaystyle \rho\mu^{2}-\left(i\mu+1\right)}\left(\xi-1\right)\nonumber \\
 & {\displaystyle c_{s2}}\equiv{\displaystyle \left(\mu-i\right)\left(\mu+i\xi\right)}\nonumber 
\end{align}

The pin blocking probability change $\frac{\Delta P_{2b}}{P_{2b}}$
in the experiment for a pin of infinitesimal width $\Delta x$ is 

\begin{equation}
\frac{\Delta P_{2b}}{P_{2b}}=-\frac{1}{P_{2b}}\frac{\delta P_{2b}}{\delta\chi(x)}\Delta x=\left(-\frac{\sqrt{-\alpha_{\chi}}}{\sqrt{\pi}}\exp(\alpha_{\chi}\left(x-x_{c}\right)^{2})+c.c.\right)\Delta x\label{dPoP}
\end{equation}

Assuming the pin has a Gaussian profile $\Delta\chi(x)=-\exp(-\frac{1}{2\sigma_{\chi}^{2}}\left(x-x_{p}\right)^{2})$
, we have

\begin{align}
 & \frac{\Delta P_{2b}}{P_{2b}}=\frac{1}{P_{2b}}\int\frac{\delta P_{2b}}{\delta\chi(x)}\Delta\chi(x)dx\\
 & =-\frac{\sqrt{-\alpha_{\chi}}}{\sqrt{\pi}}\int\exp\left(\alpha_{\chi}\left(x-x_{c}\right)^{2}-\frac{1}{2\sigma_{\chi}^{2}}\left(x-x_{p}\right)^{2}\right)dx+c.c.\nonumber 
\end{align}

After the integration, we denote the pin center $x_{p}$ as $x$,
the pin blocking probability change is,

\begin{align}
 & \frac{\Delta P_{2b}}{P_{2b}}=-\left(\frac{\alpha_{\chi}}{\alpha_{\chi}-\frac{1}{2\sigma_{\chi}^{2}}}\right)^{\frac{1}{2}}\exp\left(-\frac{\left(\frac{1}{\sigma_{\chi}^{2}}x-2\alpha_{\chi}x_{c}\right)^{2}}{4\left(\alpha_{\chi}-\frac{1}{2\sigma_{\chi}^{2}}\right)}+\alpha_{\chi}x_{c}^{2}-\frac{1}{2\sigma_{\chi}^{2}}x^{2}\right)+c.c.\label{pin_blocking_dp}
\end{align}

\subsection*{4.4 Compare the plot of $\frac{\delta P_{2b}}{\delta\chi(x,z)}$
with $\text{Re}(\psi(x,z)$), $|\psi(x,z)|^{2}$, centroid $x_{c0}$
and fringe spacing $x_{\pi}$, peak $x_{max}$ and width $\sigma_{w}$
of the envelope of $\frac{\delta P_{2b}}{\delta\chi(x,z)}$ }

For the example of setup with $s_{1}=0.0$, $\sigma_{1}=50\mu m$,
$\sigma_{2}=4\mu m$, $z_{2}=2m$, we plot $\frac{\delta P_{2b}}{\delta\chi(x,z)}$
for several pin positions $z$ compared with $\text{Re}(\psi(x,z)$),
$|\psi(x,z)|^{2}$ in Fig. 5. Fig. 5(a,b,c) is for $s_{2}=3$ mm,
$z=0.7,1.7,1.98$ m. The example of $z=1.95$m, $s_{2}=3$ mm is given
in Fig. 2(a). $\frac{\delta P_{2b}}{\delta\chi(x,z)}$ vs. $x,z$
in color scale for $s_{2}=3mm$ is given in Fig. 2(b) in the introduction. 

\begin{figure*}
\includegraphics[viewport=-14.4bp 0bp 799.2bp 432bp,clip,scale=0.32]{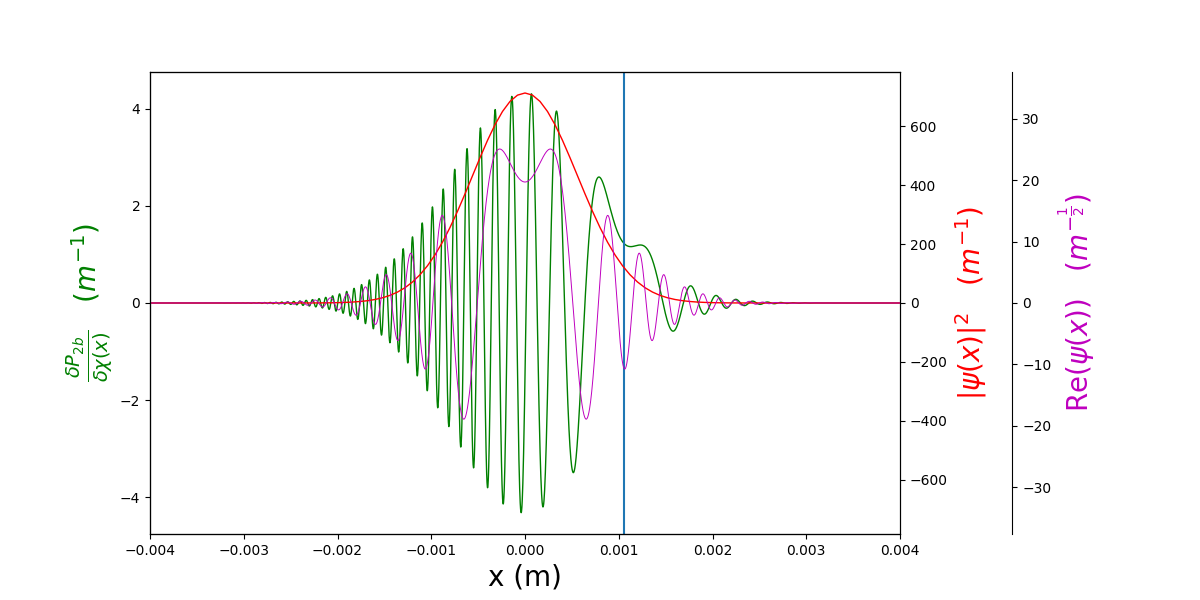}\includegraphics[viewport=43.2bp 0bp 864bp 432bp,clip,scale=0.32]{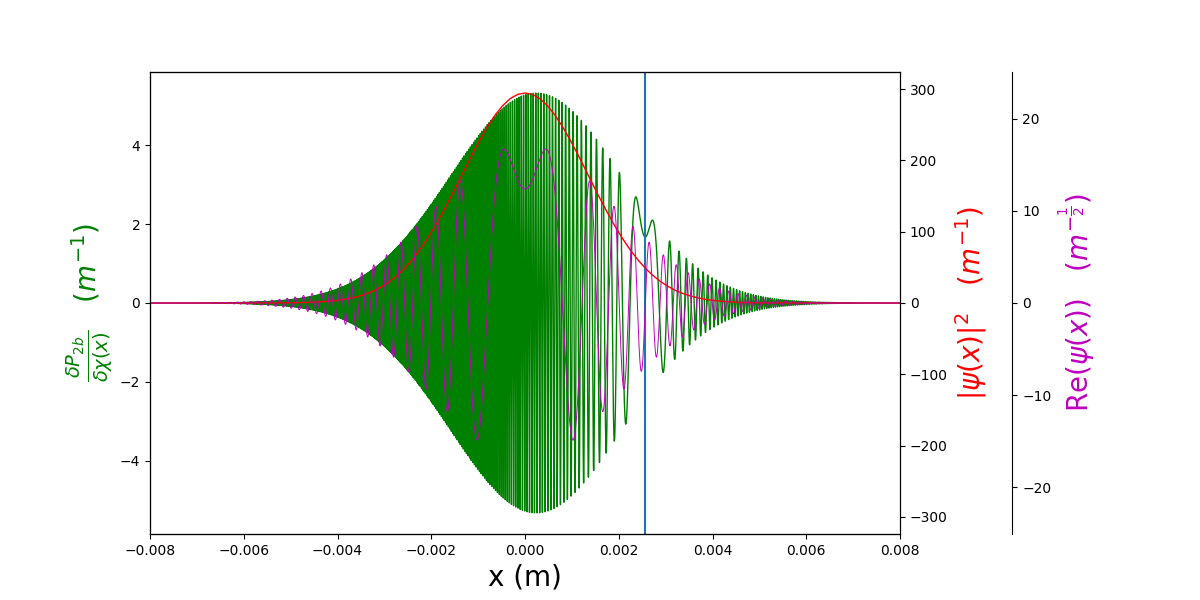}\vspace{-1.0em}

\includegraphics[viewport=36bp 0bp 792bp 432bp,clip,scale=0.35]{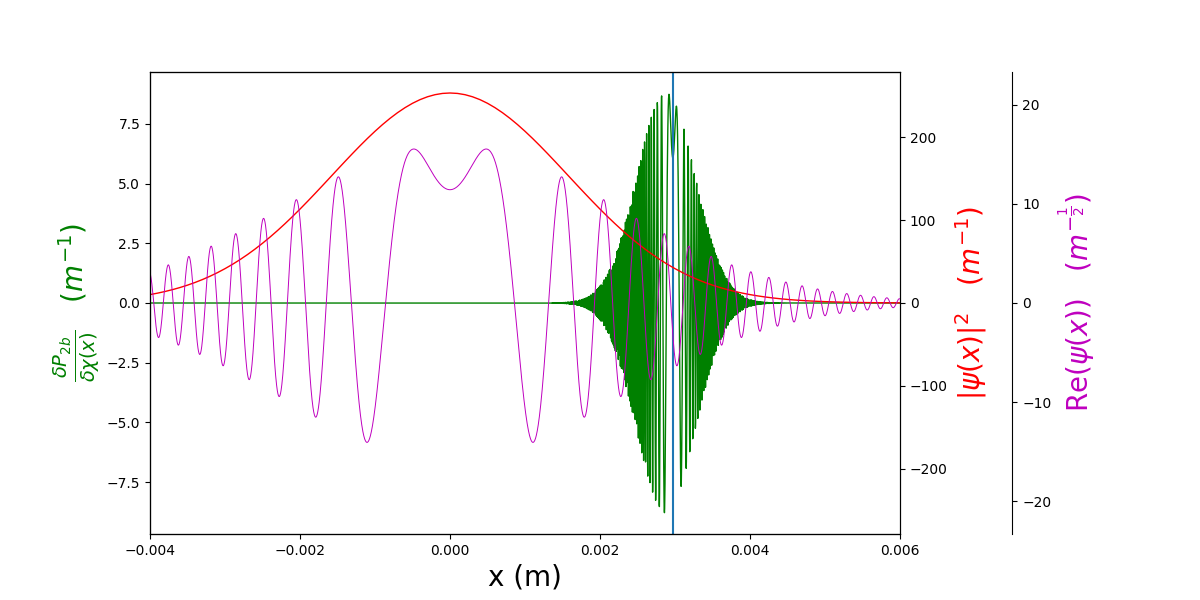}\vspace{0.0em}\includegraphics[viewport=7.20312bp 0bp 461bp 346bp,clip,scale=0.4]{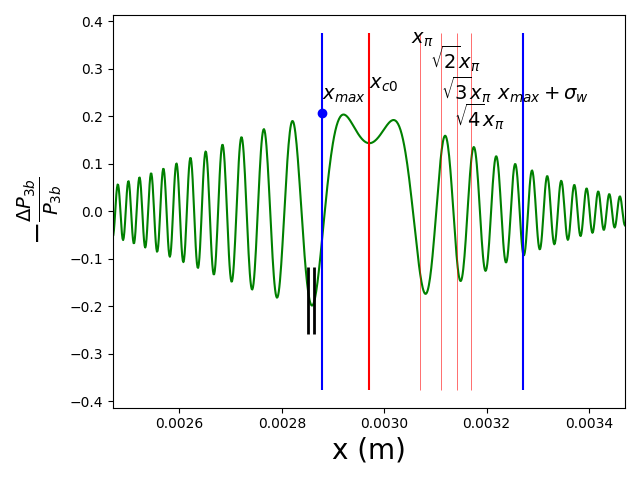}\vspace{-1.0em}

\caption{Compare diffraction pattern $|\psi(x,z)|^{2}$,$\text{Re}(\psi(x,z)$),
perturbative function $\frac{\delta P_{2b}}{\delta\chi(x)}$ of Eq.(\ref{perturbative_function})
(a)$s_{2}$=$3mm$, $z=0.7m$, (b) $s_{2}$=$3mm$, $z=1.7m$ (c)
$s_{2}$=$3mm$, $z=1.98m$. $\ensuremath{s_{1}=0.0}$, $\text{\ensuremath{\ensuremath{\sigma_{1}=50\mu m}}, \ensuremath{\sigma_{2}=4\mu m}}$,
$z_{2}=2m$. The blue vertical line is the centroid $x_{c0}$ of the
wave packet. (d) The pin blocking probability change $\frac{\Delta P_{2b}}{P_{2b}}$,
parameters are the same as (c) for region near $x_{c0}=2.97mm$, pin
width $\Delta x=10\mu m$ , indicates location of $x_{max}$, $x_{max}+\sigma_{w}$,
and $x_{c0},$$x_{c0}\pm x_{\pi},$$x_{c0}\pm\sqrt{2}x_{\pi},$$x_{c0}\pm\sqrt{3}x_{\pi},$$x_{c0}\pm\sqrt{4}x_{\pi},$with
$x_{\pi}\approx99.5\mu m$. $\sqrt{4}x_{\pi}-\sqrt{3}x_{\pi}=26.7\mu m$.
The two black vertical lines indicate the pin width.}
\end{figure*}

To be able to directly compare with the experiment, in Fig. 5(d),
we show the pin blocking probability change $\frac{\Delta P_{2b}}{P_{2b}}$
by Eq.(\ref{pin_blocking_dp}) for the same parameters in Fig. 5(c)
for the region near $x_{c0}=2.97mm$ with more details than Fig. 5(c),
and for a pin width $\sqrt{2\pi}\sigma_{\chi}=10\mu m$. Notice that
we choose the vertical axis as $-\frac{\Delta P_{2b}}{P_{2b}}$ to
compare with $\frac{\delta P_{2b}}{\delta\chi(x,z)}$ in Fig. 5(c).
For a hard-edge pin width of $\Delta x=\sqrt{2\pi}\sigma_{\chi}$,
$\frac{\Delta P_{2b}}{P_{2b}}$ would be the same when $\Delta x$
is much smaller than the fringe spacing. 

When pin width increased to $\sqrt{2\pi}\sigma_{\chi}=30\mu m$, $\frac{\Delta P_{2b}}{P_{2b}}$
amplitude increased to about 0.5 near $x_{c0}$, while far away from
$x_{c0}$, it is tiny, as shown in Fig. 6.

The blue vertical line is the centroid of the wave packet. Fig. 5
and Fig. 2 show that when $s_{2}=3mm$, $\frac{\delta P_{2b}}{\delta\chi(x,z)}$
is a wave packet with its centroid moving from $x=0,z=0$ to $x=3mm,z=2m$.
The wave packet has both amplitude and phase information. The centroid
is where the phase is stationary, i.e., it is where the phase $\text{Im}(\alpha_{\chi}\left(x-x_{c}\right)^{2})$
as the function of $x$ has its derivative $\frac{\partial}{\partial x}\text{Im}(\alpha_{\chi}\left(x-x_{c}\right)^{2})=0$.
Solving this equation, we find the stationary point is 

\begin{equation}
x_{c0}=\frac{\alpha_{\chi i}x_{cr}+\alpha_{\chi r}x_{ci}}{\alpha_{\chi i}}=x_{cr}+\frac{\alpha_{\chi r}}{\alpha_{\chi i}}x_{ci}\label{xc0}
\end{equation}

Where $\alpha_{\chi r},\alpha_{\chi i},x_{cr},x_{\chi i}$ are the
real part and imaginary part of $\alpha_{\chi},x_{c}$ such that $\alpha_{\chi}=\alpha_{\chi r}+i\alpha_{\chi i}$,$x_{c}=x_{cr}+ix_{\chi i}$
as calculated from Eq.(\ref{alphachi-2},\ref{xc-2}).

The $x$-dependent part of the phase of $\frac{\delta P_{2b}}{\delta\chi(x,z)}$
is given by $\theta=\text{Im\ensuremath{\left({\displaystyle \alpha_{\chi}\left(x-x_{c}\right)^{2}}\right)}}$,
$\theta$ changes quadratically when $x$ is away from the centroid
$x_{c0}$, the further away from the $x_{c0}$, the faster it varies.
To calculate the fringe spacing, we assume the points are spaced with
a phase difference $\Delta\theta=n\pi$ from the centroid $x_{c0}$

\begin{align}
 & \Delta\theta=\text{\text{Im}}\left(\left({\displaystyle \alpha_{\chi}\left(x_{c0}+\Delta x-x_{c}\right)^{2}}\right)-\left({\displaystyle \alpha_{\chi}\left(x_{c0}-x_{c}\right)^{2}}\right)\right)=n\pi\label{xpi}
\end{align}

From this equation, we find $\Delta x=\pm x_{\pi}\sqrt{n}$, $x_{\pi}=\sqrt{\frac{\pi}{\alpha_{\chi i}}}$.
For the example in Fig. 5(d), $x_{c0}\approx2.970mm$ $\alpha_{\chi}\approx{\displaystyle \left(-0.03268+3.1729i\right)\times10^{8}}m^{-2}$,
$x_{\pi}\approx99.5\mu m$. We use the thin red lines to indicate
the fringe spacing; they are not at peaks of the fringes because the
phase of $x_{c0}$ is not zero.

Using the same method, we find the position of the peak of the envelope
of $\frac{\delta P_{2b}}{\delta\chi(x,z)}$, i.e., $|P_{2b}\frac{\sqrt{-\alpha_{\chi}}}{\sqrt{\pi}}\exp(\alpha_{\chi}\left(x-x_{c}\right)^{2})|$
as $x_{max}=\frac{\alpha_{\chi r}x_{cr}-\alpha_{\chi i}x_{ci}}{\alpha_{\chi r}}=2.878mm$
and its width $\sigma_{w}^{2}=-\ensuremath{\frac{1}{2\alpha_{\chi r}}}$
with $\sigma_{w}=392\mu m$ indicated by the blue lines in Fig. 5(d).
We see that $x_{c0}$ is not at the peak of the envelope of the green
curve of $\frac{\delta P_{2b}}{\delta\chi(x,z)}$, and it is further
away from the peak of the wave function $\psi(x,z)$.

Fig. 2(b) and Fig. 5 show that $\frac{\delta P_{2b}}{\delta\chi(x,z)}$
represents a wave packet moving towards the exit slit, and when close
to the exit point, the width $\sigma_{w}$ becomes very small, and
the fringe spacing $x_{\pi}$ rapidly becomes very narrow and hardly
visible, and hence becomes more difficult to measure. 

\begin{figure*}
\includegraphics[viewport=-14.40625bp 0bp 799.547bp 432.5bp,clip,scale=0.4]{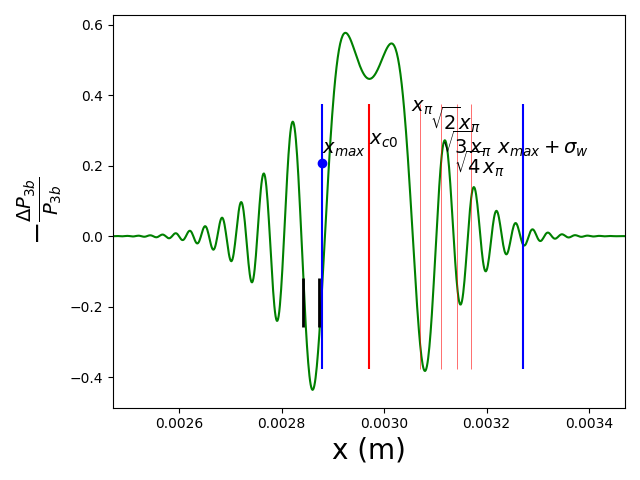}

\caption{The pin blocking probability change $\frac{\Delta P_{2b}}{P_{2b}}$,
parameters are the same as Fig. 5(d), with pin width increased to
$\sqrt{2\pi}\sigma_{\chi}=30\mu m$. The two black vertical lines
indicate the pin width.}
\end{figure*}
For a comparison, the plot of $\text{Re}(\psi(x,z))$ and $|\psi(x,z)|^{2}$
are entirely different; the width increases as $z$ increases until
at the exit slit, where it is widespread. $\text{Re}(\psi(x,z))$
has fringes, but we can only measure $|\psi(x,z)|^{2}$. 

To exhibit the fringe width $x_{\pi}(z)$ dependence on $z$, we plot
$x_{c0}(z)\pm\frac{1}{2}x_{\pi}(z)$ as a function of $z$ for slit
2 position $s_{2}=0,1,2,3$ mm, respectively, in Fig. 3 in the introduction
to compare with the width of $|\psi(x,z)|^{2}$.

We consider the well-known wave function $\psi(x,z)$ to represent
a cylindrical wave starting from the entrance slit and continuously
spreading out until it finally arrives at the screen of the exit slit.
Notice that the plots of $\text{Re}(\psi(x,z)$) and $|\psi(x,z)|^{2}$
in Fig. 5 and Fig. 2, 3 do not show the cylindrical waveform explicitly
because the phase represented in Fig. 5 is only the phase of $\psi(x,z)$
in $\phi(x,z,t)=\exp\left(i(kx-\omega_{k}t)\right)\psi(x,z)$ as defined
in Section 3.1; it does not include the longitudinal phase $(kx-\omega_{k}t)$.
The cylindrical waveform is implicitly exhibited only when we examine
the waveform of $\text{Re}(\psi(x,z))$ and see it spreading out as
$z$ increases from $z=0.7m$ to 1.7m and 1.98m. It would not be so
easy to show the cylindrical waveform because, in a $(z,x)$ 2D plot,
the phase $(kx-\omega_{k}t)$ variation will dominate over the phase
variation of the slow-varying amplitude and phase function $\psi(x,z)$
for a case where the paraxial approximation is valid. The curvature
of the cylindrical wavefront becomes visible only when $x$ is the
same order as $z$. But we only consider those regions where $|x|\ll z$
in the paraxial approximation.

We discuss this cylindrical waveform here mainly because we would
like to compare it to $\frac{\delta P_{2b}}{\delta\chi(x,z)}$ shown
in Fig. 2, Fig. 3, and Fig. 5, where it exhibits an entirely different
picture of wave packets emitted from the entrance slit into various
directions and striking on the exit screen at different points. The
wave packet width is very narrow at the entrance slit, grows wider
and reaches maximum width, then becomes narrower again until it converges
into the exit slit, forming a spindle-shaped region.

\subsection*{4.5 The small slit size limit }

It would be interesting to study the small slit size limit because
in this limit whether the profile of the entrance slit and exit slit
are Gaussian or hard edge is no longer important. But when we try
to work out the calculation replacing the Gaussian functions $f_{1},f_{2}$
by $\delta$-functions, we have some difficulties because $P_{2b}$
and $\frac{\delta P_{2b}}{\delta\chi(x,z)}$ become zero. Then, we
first work out the expression for $\frac{\delta P_{2b}}{\delta\chi(x,z)}$,
and now we take this limit. The following will show that the ratio
$\rho=\frac{\sigma_{2}^{2}}{\sigma_{1}^{2}}$ is relevant and can
be derived from the measurement of $\frac{\delta P_{2b}}{\delta\chi(x,z)}$. 

The example we have for Fig. 2-4 all have small $\sigma_{1}$ and
$\sigma_{2}$, where $\mu=\frac{\sigma_{1}^{2}}{d_{1}}=\frac{4\pi\sigma_{1}^{2}}{\lambda L}\ll1$,
If $z$ is not very close to 0 or $L$ so $\xi=\frac{z}{L}$ and $1-\xi=\frac{L-z}{L}$
is too far from order of magnitude 1, we can take the approximation
in Eq.(\ref{alphachi-2}) 

\begin{align}
 & {\displaystyle \alpha_{\chi}=\alpha_{\chi r}+i\alpha_{\chi i}\approx-\frac{\mu^{2}}{2\sigma_{1}^{2}}\frac{{\displaystyle \left(\rho\xi^{2}+2(\xi-1)^{2}\right)}}{{\displaystyle 4\xi^{2}\left(\xi-1\right)}^{2}}-i\frac{\mu}{4\sigma_{1}^{2}}\frac{{\displaystyle 1}}{{\displaystyle \xi\left(\xi-1\right)}}}\label{alphachi_lim}
\end{align}

We find the phase shift $\pi$ width $x_{\pi}=\sqrt{\frac{\pi}{\alpha_{\chi i}}}\approx\sqrt{\lambda L\xi\left(1-\xi\right)}$.
This width is shown in Fig. 7 as the spindle-shaped region of $\frac{\delta P_{2b}}{\delta\chi(x,z)}$
reaches maximum width at $z=L/2$ in the middle of slits 1 and 2.
The wave packet envelope width is also given in Fig. 7 as $\sigma_{w}\approx\frac{\lambda L}{2\sqrt{2}\pi\sigma_{1}}\ensuremath{\sqrt{\frac{\xi^{2}{\displaystyle \left(\xi-1\right)}^{2}}{{\displaystyle \frac{\rho}{2}\xi^{2}+}{\displaystyle \left(\xi-1\right)}^{2}}}}$.
The maximum of $\sigma_{w}$ is the root of a cubic equation ${\displaystyle \rho\xi^{3}+2\xi^{3}-6\xi^{2}+6\xi-2}=0$
, indicated by a vertical blue line. Fig. 7 shows the width $x_{\pi}$
and $\sigma_{w}$ of the wave packet $\frac{\delta P_{2b}}{\delta\chi(x,z)}$
in the small slit size limit agrees with the formula in Eq.(\ref{alphachi-2})
well. 

The result shows the relative position $\xi=z/L$ of maximum width
$\sigma_{w}$ depends on $\rho=\frac{\sigma_{2}^{2}}{\sigma_{1}^{2}}$,
so it is not in the middle like the position of maximum $x_{\pi}$
always in the middle.

However, since $P_{2b}=\frac{1}{2}\int_{-\infty}^{\infty}dx\frac{\delta P_{2b}}{\delta\chi(x,z)}$,
contribution to $P_{2b}$ mainly comes from near $x_{c0}$. Far from
$x_{c0}$, $\frac{\delta P_{2b}}{\delta\chi(x,z)}$ oscillates fast,
and the positive and negative peaks cancel each other. Hence, the
width $x_{\pi}$ is more relevant to the total probability $P_{2b}$
than the envelope width $\sigma_{w}$. Thus, the spindle-shaped region
visible in Fig. 2(b) and Fig. 3  is mainly determined by the spindle-shaped
curve of $x_{\pi}$ in Fig. 7. When one examines Fig. 2(b) in detail,
it is also possible to observe the shape of the curve $\sigma_{w}$
in Fig. 2(b).

\begin{figure*}
\includegraphics[scale=0.5]{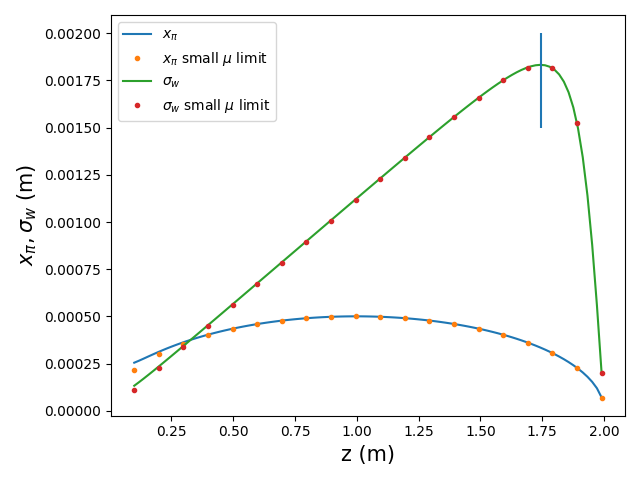}\vspace{-1.0em}

\caption{width $x_{\pi},\sigma_{w}$ calculated from Eq.(\ref{alphachi-2})
compared with the limit $\sigma_{1},\sigma_{2}\rightarrow0$. The
maximum width $x_{\pi}$ of phase shift $\pi$ is in the middle point
$z=1m$, and the maximum of the wave packet envelope is at $z=1.743m$
indicated by the blue vertical line corresponding to the solution
of the cubic equation $\xi=z/L=0.872$.}
\vspace{-1.0em}
\end{figure*}

\subsection*{4.6 Counting number and relation to measurement error bar}

Take a point in Fig. 5(d) at $x=x_{p}=2.859mm$ as the pin position,
and take a width $\Delta x=10\mu$ to represent the blocking pin indicated
by the two vertical lines in the figure. $\frac{\delta P_{2b}}{\delta\chi(x,z)}\approx{\displaystyle -8.58m^{-1}\pm0.18}$
within this $10\mu m$. $P_{2b}=0.0004258$ is already calculated
in Section 3.3 for the same set of parameters as Fig. 5(d). The contribution
of $\Delta x$ to the final probability is $\frac{\Delta P_{2b}}{P_{2b}}=\frac{1}{P_{2b}}\frac{\delta P_{2b}}{\delta\chi(x,z)}\Delta x\approx\frac{{\displaystyle -8.58}/m}{0.0004258}\times10\mu m=-0.20$.
If we block a line of width $\Delta x$ that passes through x (this
line is in parallel in the y-direction) such that immediately behind
this line $\psi(x)=0$, then the probability of passing through the
end slit 3 will decrease by a factor of $-0.20$, i.e., the probability
will increase by a factor of $1.20$. 

To get sufficient accuracy, we use the Poisson distribution formula
$P(k)={\displaystyle \frac{n^{k}e^{-n}}{k!}}$, its RMS is $\sqrt{n}$
where $n$ is the mean rate of events during a fixed interval.

If the recorded detection count $n=4\times10^{4}$, then the RMS fluctuation
is $\Delta n=\sqrt{n}=200$, $\frac{\Delta n}{n}=0.005$, and the
probability of detection when normalized as $1$ without blocking
is increased from $1$ to $1.20\pm0.005$. Thus the error of the measurement
$\frac{\delta P_{2b}}{\delta\chi(x,z)}$ is $0.005/0.2=0.025$. The
error due to the variation of $\frac{\delta P_{2b}}{\delta\chi(x,z)}$
within the pin width is $0.18/8.58=0.02$. There is a trade-off between
these two different error types when choosing the pin width.

We restrict our discussion to the fundamental principles of the experiment,
where we have assumed a detection quantum efficiency of 100\%, omitting
further details. In reality, the lower detector quantum efficiency
may correspond to taking a smaller $\sigma_{2}$, while the stability
of the system and the background noise may be more important for the
precision of the measurement.

\section*{5. Wave packet propagation in an interference experiment}

To demonstrate the phase information inherent in $\frac{\delta}{\delta\chi(x,z)}P_{2b}$,
we study an interference experiment as shown in Fig. 4. We replace
the single entrance slit 1 in Fig. 1 with two slits. To keep coherence
between the two slits, the wavefront is assumed to arrive at the two
slits in phase. The initial wave function at the two slits is set
the same except with a transverse displacement of $2s_{1}$,

\begin{align}
 & \psi_{1}(x_{1})=\frac{\left(\psi_{1+}(x_{1})+\psi_{1-}(x_{1})\right)}{\sqrt{2}}\label{psi1}\\
 & \psi_{1\pm}(x_{1})=f_{1}(x_{1}\mp s_{1})=\left(\frac{1}{2\pi\sigma_{1}^{2}}\right)^{\frac{1}{4}}\exp\left(-\frac{1}{4\sigma_{1}^{2}}(x_{1}\mp s_{1})^{2}\right)\nonumber 
\end{align}

They centered at $s_{1}=1$ mm and $s_{1}=-1$ mm respectively, and
are in phase. Their separation $2s_{1}\gg\sigma_{1}$, so their overlap
is negligible, and hence the normalization $P_{1}=\int dx_{1}|\psi_{1}(x_{1},z_{1}=0)|^{2}=1$
is satisfied. 

To calculate the perturbative function $\frac{\delta}{\delta\chi(x,z)}P_{2b}$,
we first substitute this into Eq.(\ref{perturbative_function}) to
calculate the two functions $\phi_{b}(x),\psi(x)$, we get

\begin{align}
 & \frac{\delta}{\delta\chi(x)}P_{2b}=\phi_{b}(x)\psi(x)+c.c.\nonumber \\
 & \psi(x)\equiv\int dx_{1}G(x,x_{1};t)\psi_{1}(x_{1})=\frac{1}{\sqrt{2}}\int dx_{1}G(x,x_{1};t)\left(\psi_{1+}(x_{1})+\psi_{1-}(x_{1})\right)\equiv\frac{1}{\sqrt{2}}\left(\psi_{+}(x)+\psi_{-}(x)\right)\\
 & \psi_{2b+}(x_{2})=f_{2}(x_{2})\int dx_{1}G(x_{2},x_{1};t_{1},0)\psi_{1+}(x_{1})\nonumber \\
 & \psi_{2b-}(x_{2})=f_{2}(x_{2})\int dx_{1}G(x_{2},x_{1};t_{1},0)\psi_{1-}(x_{1})\nonumber \\
 & \phi_{b}(x)\equiv\int dx_{2}\left(\psi_{2b}^{*}(x_{2},t_{1})f_{2}(x_{2})G(x_{2},x;t_{L}-t)\right)\nonumber \\
 & \equiv\frac{1}{\sqrt{2}}\int dx_{2}\left\{ \left(\psi_{2b+}^{*}(x_{2},t_{1})+\psi_{2b-}^{*}(x_{2},t_{1})\right)f_{2}(x_{2})G(x_{2},x;t_{L}-t)\right\} \equiv\frac{1}{\sqrt{2}}\left(\phi_{b+}(x)+\phi_{b-}(x)\right)\nonumber 
\end{align}

We can obtain each of the four terms $\psi_{+}(x),\psi_{-}(x),\phi_{b+}(x),\phi_{b-}(x)$
in a more explicit form in Appendix IV.1 by replacing $s_{1}$ by
either $s_{1}$ or $-s_{1}$ in the expression of $\phi_{b}(x)$ of
Eq.(\ref{lnphib}) and separately the expression of $\psi(x)$ of
Eq.(\ref{lnphi2a}). Then we get the perturbative function between
the entrance and the slit 2

\begin{align}
 & \frac{\delta}{\delta\chi(x)}P_{2b}=\frac{\left(\phi_{b+}(x)+\phi_{b-}(x)\right)}{\sqrt{2}}\frac{\left(\psi_{+}(x)+\psi_{-}(x)\right)}{\sqrt{2}}+c.c.\label{interference2}\\
 & =\frac{1}{2}\left(\phi_{b+}(x)\psi_{+}(x)+\phi_{b-}(x)\psi_{-}(x)+\phi_{b+}(x)\psi_{-}(x)+\phi_{b-}(x)\psi_{+}(x)+c.c.\right)\nonumber 
\end{align}

So the interference pattern $P_{2b}(s_{2})$ is obtained by the integration
in Eq.(\ref{PdPrelation}), using the Gaussian integral Eq.(\ref{gaussian_integral})
in Appendix II.1,

\begin{align}
 & P_{2b}=\frac{1}{2}\int_{-\infty}^{\infty}dx\frac{\delta}{\delta\chi(x)}P_{2b}=\int_{-\infty}^{\infty}dq\frac{1}{4}\left(\phi_{b+}(x)\psi_{+}(x)+\phi_{b-}(x)\psi_{-}(x)+\phi_{b+}(x)\psi_{-}(x)+\phi_{b-}(x)\psi_{+}(x)+c.c.\right)\nonumber \\
 & =\int_{-\infty}^{\infty}dq\frac{1}{2}\text{Re}\left(\phi_{b+}(x)\psi_{+}(x)+\phi_{b-}(x)\psi_{-}(x)+\phi_{b+}(x)\psi_{-}(x)+\phi_{b-}(x)\psi_{+}(x)\right)\label{P3b}
\end{align}

For the same set of parameters as in Fig. 2, i.e., photon wavelength
$\lambda=0.5\mu m$, $z_{3}=2m,\text{\ensuremath{\sigma_{1}=50\mu m}, \ensuremath{\sigma_{3}=4\mu m}}$
except that $s_{1}=0$ is replaced by $\pm s_{1}=\pm1$ mm for the
double slit experiment, Fig. 4(a) shows $P_{2b}(s_{2})=\frac{1}{2}\intop_{-\infty}^{\infty}dx\frac{\delta P_{2b}}{\delta\chi(x,z)}$
calculated at $z=1.9m$ as the well-known interference pattern. However,
as discussed above, the plot is independent of $z$: it is the same
for any $0<z<2m$. In Fig.4(b), we choose $s_{2}=0$, where $P_{2b}=0.004122$
reaches the peak. To verify this result, we use the method in Eq.(\ref{P3bo})
to directly calculate $P_{2b}=0.002061$ with only one of the slits
open. So the interference peak is twice the probability of a single
slit. It is only twice, not 4 times because the normalization of $\psi_{1}(x_{1})$
requires that it is reduced by $\sqrt{2}$ in Eq.(\ref{psi1}). For
this peak point shown as the red dot in Fig. 4(a), the perturbative
function $\frac{\delta P_{2b}}{\delta\chi(x,z)}$ waveform is shown
in Fig. 4(b). There are two spindle-shaped regions symmetrically oriented
where $\frac{\delta P_{2b}}{\delta\chi(x,z)}>0$ varies slowly, and
both have a color between yellow to deep red and are in phase. In
Fig. 4(c), we choose $s_{2}=-0.25mm$, where $P_{2b}(s_{2})=\frac{1}{2}\intop_{-\infty}^{\infty}dx\frac{\delta P_{2b}}{\delta\chi(x,z)}=1.249\times10^{-5}$
reaches the minimum shown in Fig. 4(a) as the green dot; the two spindle-shaped
regions have different colors. In the upper plane, the region is colored
red. The lower region is colored blue, with a phase difference of
$\pi$. Because of the finite size of $\sigma_{2}$, the minimum $P_{2b}=1.249\times10^{-5}$
is not zero. In this case, the positive peaks of $\frac{\delta P_{2b}}{\delta\chi(x,z)}$
in the integration $P_{2b}=\frac{1}{2}\intop_{-\infty}^{\infty}dx\frac{\delta P_{2b}}{\delta\chi(x,z)}$
almost cancel the negative peaks over the $x-z$ plane. Notice that
even though $\frac{\delta P_{2b}}{\delta\chi(x,z)}$ depends on $z$,
the integral is independent of $z$. This independence further confirms
what we emphasized in Section 4.2 about Eq.(\ref{PdPrelation}) that
the probability $P_{2b}$ is already determined by sampling from a
distribution right in the beginning at the entrance slit 1 with $z=0_{+}$,
even though it is measured at the exit slit after the wave packet
(the particle) passes through the whole system.

\section*{6. Discussion on a Few Questions about the Wave Function Collapse}

We discuss our understanding regarding a few questions suggested by
the analysis above.

The discussion of Eq.(\ref{PdPrelation}) in Section 4.2 shows that
the detection probability $P_{2b}=\frac{1}{2}\int_{-\infty}^{\infty}dx\frac{\delta P_{2b}}{\delta\chi(x,z)}$
is independent of $z$, meaning immediately after passing through
the entrance slit, the integral of $\frac{1}{2}\int_{-\infty}^{\infty}dx\frac{\delta P_{2b}}{\delta\chi(x)}$
has already been determined as $P_{2b}$ at the exit. 

One might have a few questions. First, why in the very beginning,
right after passing the slit 1, the probability is already determined
long before the particles arrive at the exit slit. However, this does
not violate the causality because the initial wave function $\psi_{1}(x_{1},z_{1})$
only gives a distribution of states based on the Born rule, and the
final result with probability $P_{2b}$ is only one state in the distribution. 

Second, another question arises when we notice that the function $\frac{\delta P_{2b}}{\delta\chi(x,z=0_{+})}$
depends on $z_{2}=L$, so if in a thought experiment, we move the
end slit longitudinally to a different $L$ after the particle has
passed the entrance slit, in principle the function $\frac{\delta P_{2b}}{\delta\chi(x,z=0_{+})}$
would have changed too even before the particle arrives at the exit
slit. The question would be how the position change of the detector
at the end can influence the function $\frac{\delta P_{2b}}{\delta\chi(x,z=0_{+})}$
determined at the beginning. However, this also does not violate causality
because the wave function evolution is determined by the unitary transform
given by the Schr\textcyr{\"\cyro}dinger equation and because of the
time-reverse symmetry of the equation. We may consider the set \{$s_{2n}$\}
of the detectors arranged to extend over the exit screen we mentioned
before. When finally the particle is detected at $s_{2n}$, its state
can only be a $\delta$-function at $s_{2n}$, initially, the perturbative
function must be $\frac{\delta P_{2b}(s_{2n})}{\delta\chi(x,z=0_{+})}$
because only the state represented by $\frac{\delta P_{2b}(s_{2n})}{\delta\chi(x,z=0_{+})}$
can evolve into the point $s_{2n}$ at the detector. When we prepare
the initial state as $\psi_{1}(x_{1},z_{1})$ by sending particles
into the entrance slit, it must have immediately become a distribution
if we set up the set of detectors $\{s_{2n}\}$ at $t_{2}=\frac{z_{2}}{v}$. 

Another way to understand this question is to point out the relation
between the eigenvectors of the position operator in the Heisenberg
picture and their relation to $\frac{\delta P_{2b}(s_{2n})}{\delta\chi(x,z=0_{+})}$.
According to the discussion following Eq.(\ref{perturbative_function}),
when $\sigma_{1},\sigma_{2}$ approach zero, $\frac{\delta P_{2b}(s_{2n})}{\delta\chi(x,z=0_{+})}$
is proportional to $G(x_{2},x;t_{2}-t_{1})G(x,x_{1};t-t_{1})$, and
approaches $G(x_{2},x_{1};t_{2}-t_{1})$ as $x$ approaches $x_{1}$.
According to Eq.(\ref{greens_function}) in Appendix I, $G(x_{2},x_{1};t_{2}-t_{1})=u_{x_{2}}^{*}(x_{1},t_{2}-t_{1})$.
In other words, $\frac{\delta P_{2b}(s_{2n})}{\delta\chi(x,z=0_{+})}$
is the Feynman propagator from $0_{+}$ to $z_{2}$. 

In the Heisenberg picture, this becomes clearer, as we discussed in
Appendix I. We denote the initial position operator in the Heisenberg
picture as $X_{H}(t_{1})=X_{S}$ because it is the same as in the
Schr\textcyr{\"\cyro}dinger picture. We denote the position operator
at time $t_{2}$ as $X_{H}(t_{2})$ and denote the evolution operator
from $t_{1}$ to $t_{2}$ as $U(t_{2}),$ then $X_{H}(t_{2})=U^{\dagger}X_{S}U$.
We denote the eigenstate of $X(t_{2})$ as $u_{x_{2}}(x_{0})$, i.e.,
$X(t_{2})u_{x_{2}}(x_{0})=x_{2}u_{x_{2}}(x_{0})$. Because a state
vector in the Heisenberg picture does not change with time, the state
$u_{x_{2}}(x_{0})$ is also the initial state vector in the Schr\textcyr{\"\cyro}dinger
picture. In the Schr\textcyr{\"\cyro}dinger picture, the state $u_{x_{2}}(x_{0})$
will evolve into the eigenvector $U(t_{2})u_{x_{2}}(x_{0})$, and
this must be the eigenstate of the $X_{S},$ i.e., $X_{S}U(t_{2})u_{x_{2}}(x_{0})=x_{2}U(t_{2})u_{x_{2}}(x_{0})$.
Hence, $U(t_{2})u_{x_{2}}(x_{0})$ is the $\delta$-function at $x_{2}$.
In other words, $u_{x_{2}}(x_{0})$ is the initial wave function that
eventually evolves into the eigenstate of position operator $X_{S}$
with eigenvalue $x_{2}$ in the Schr\textcyr{\"\cyro}dinger picture.
If the detector at $s_{2n}$ detects the particle, then $x_{2}=s_{2n}$.
We use the Heisenberg picture to clarify this point here mainly because
the state vector, i.e. the wave function does not change. However,
we use the Schr\textcyr{\"\cyro}dinger picture to derive the perturbative
function through out in this paper because the main purpose is to
understand how the perturbative function evolves.

The complete orthonormal basis $\{u_{x_{2}}(x_{0})\}$ in the Schr\textcyr{\"\cyro}dinger
picture will evolve into the complete basis consisting of $\delta$-functions
at $\{x_{2}\}$. Thus, the initial state can only be one in this set
of functions $\{u_{x_{2}}(x_{0})\}$ as long as we set the detectors
at position $z_{2}=L$. This set of functions depends on $L$. If
we change $L$, the set $\{u_{x_{2}}(x_{0})\}$ will change too. There
is no causality violation here because the unitary transform by the
evolution operator $U(t_{2})$ determines this relation between the
position of detectors and $\{u_{x_{2}}(x_{0})\}$. The wave function
is a linear combination of $\delta$-functions at time $t_{2}$, and
the probability distribution follows the Born rule. When we apply
the unitary transform $U^{-1}$ to this final distribution of $\delta$-functions
to derive the initial function, the result must be the linear combination
of $\{u_{x_{2}}(x_{0})\}$, and the probability distribution must
follow the Born rule. This conclusion means the initial state cannot
be a $\delta$-function; it has to be a distribution of $\{u_{x_{2}}(x_{0})\}$.
If the experiment prepares a $\delta$-function as the initial state,
it would immediately become a distribution of $\{u_{x_{2}}(x_{0})\}$
according to the Born rule. It is just like if we view an object from
different perspectives, we will reveal other aspects of the object.
According to the uncertainty principle, we also reach the same conclusion:
$X_{H}(t_{1})$ and $X_{H}(t_{2})=X_{S}+\frac{P_{s}}{M}(t_{2}-t_{1})$
do not commute; they cannot have the same eigenvector. When the detector
$s_{2n}$ detects a particle, its state vector is an eigenstate of
$X_{H}(t_{2})$; hence, its initial state cannot be an eigenstate
of $X_{H}(t_{1}).$ Even though we prepare an eigenstate of the initial
$X_{H}(t_{1})$, it is already a distribution of $\{u_{x_{2}}(x_{0})\}$
the moment we prepare it because the detector array is not at the
origin at slit 1.

A change in the position of point 2 affects the selection of past
event data used for the measurement. The measurement in the experiment
is the probability for the initial state at time $t_{1}$ being the
eigenvector of $x_{2}(t_{2})$. If we change the way of sorting out
historical data, we will get a different set of data.

If we take $t=0$ as the start time, then there is a collapse of knowledge.
If we take $t=0_{+}$ as a starting point, the initial state is just
a distribution derived from the expansion of $\delta$-function in
terms of $\{u_{x_{2}}(x_{0})\}$ with probability determined by the
Born rule. According to the Schr�dinger\textendash HJW theorem \cite{Kirkpatrick,HJW theorem},
the pure state of a $\delta$-function is equivalent to a mixed state
expressed by a density matrix with a unitary transform.

However, the experiment's precision is always limited, so the discussion
above does not exclude the wave function collapse but only gives a
possible upper limit for the time scale $\epsilon$ of the wave function
collapse , as discussed in Section 2 following the introduction to
Section 4. Thus, $\epsilon$ provided by an experiment may offer a
limited time scale for the wave function collapse for an objective-collapse
theory \cite{objective_collapse}.

\section*{7. Summary}

We study the wave packet evolution when a single particle travels
in the free space between two slits before it arrives at the detector
behind the exit slit. We use a narrow pin to perturbatively block
the wave function in repeated experiments to count the detection rate
as a function of the pin's transverse and longitudinal position $(x,z)$. 

When we normalize the probability of entering the entrance slit as
1, the perturbative function $\frac{\delta P_{2b}}{\delta\chi(x,z)}$
measured this way is the functional derivative of the probability
$P_{2b}$ of detection. The result shows this function is a real-valued
function with phase information. The shape of the function is very
different from the well-known solution $\psi(x,z)$ of the Schr\textcyr{\"\cyro}dinger
equation with the given initial wave function $\psi_{1}(x_{1},z_{1})$.
This perturbative function $\frac{\delta P_{2b}}{\delta\chi(x,z)}$
forms a spindle-shaped region between the two slits with two pointed
ends positioned at the two slits, while $|\psi(x,z)|^{2}$ forms a
fan-shaped region with only the pointed end at the entrance slit.
While we determine $\psi(x,z)$ by specifying the initial value of
$s_{1}$ only, the perturbative function $\frac{\delta P_{2b}}{\delta\chi(x,z)}$
is determined by both the initial $s_{1}$ and final $s_{2}$, somewhat
like in the classical theory, where we determine the particle trajectory
by two points, i.e., the initial and final position. While $|\psi(x,z)|^{2}$
provides a statistical distribution of final states, the perturbative
function $\frac{\delta P_{2b}}{\delta\chi(x,z)}$ provides a specific,
definite, completely reproducible result once $s_{2}$ is set up in
an experiment.

The integral $\frac{1}{2}\intop_{-\infty}^{\infty}dx\frac{\delta P_{2b}}{\delta\chi(x,z)}$
at any given $z$ between the two slits provides the detection probability
$P_{2b}$ no matter where $z$ is, even for $z=0_{+}$, i.e., right
at the beginning. This result shows the probability of detection at
a given point $s_{2}$ is determined long before the particle arrives
at the detector. This outcome, in turn, means that once the experiment
confirms the prediction of $\frac{\delta P_{2b}}{\delta\chi(x,z)}$,
it provides a picture of the wave packet evolution different from
that of a cylindrical wavefront of propagation that starts from the
source slit and then continues to widespread at the exit slit. The
picture provided by $\frac{\delta P_{2b}}{\delta\chi(x,z)}$ is that
the source emits spindle-shaped wave packets into various directions
to arrive at different detectors at various positions indexed as $s_{2n}$,
each with a probability $P_{2b}(s_{2n})$. If the width $\sigma_{2}$
of slit 2 is sufficiently small, in a hypothetical experiment, a set
of slits $s_{2n}$ spaced by $\sqrt{2\pi}\sigma_{2}$ extending over
the exit screen would have a combined detection probability of one
for a set of idealized detectors with 100\% quantum efficiency. This
probability of one means the probability amplitude $\psi(x,z)$ represents
the distribution of an ensemble of single-particle wave packets. While
$\frac{\delta P_{2b}}{\delta\chi(x,z)}$ represents one specific state
in the ensemble. 

Thus, one can separate the wave function collapse process into two:
First, from the ensemble of states, one state is sampled; second,
in the specific sampled state, the particle wavepacket evolves according
to the function $\frac{\delta P_{2b}}{\delta\chi(x,z)}$ starting
from the source point, growing to a maximum in the middle of the slit
1 and 2, then converging to the exit slit 2. The first step is more
like a conceptual change of our knowledge from uncertainty to certainty
rather than a continuous process like the one in the second step,
where a wave packet collapses into a point embodied as a particle.

However, the experiment's precision is always limited. It gives a
possible limit for the time scale $\epsilon$ of single particle wave
function collapse, to be determined by the experiment, as discussed
in Section 2 and Section 6. Thus, $\epsilon$ provided by an experiment
may hint at selecting some parameters for an objective-collapse theory
\cite{objective_collapse}.

We find the explicit expression $\frac{\delta P_{2b}}{\delta\chi(x,z)}=P_{2b}\frac{\sqrt{-\alpha_{\chi}}}{\sqrt{\pi}}\exp(\alpha_{\chi}\left(x-x_{c}\right)^{2})+c.c.$
, and we derive the expression for $P_{2b},$ $\alpha_{\chi}$, and
$x_{c}$. Then, from these, we calculated the maximum phase $\pi$
shift width $x_{\pi}$ , and the envelope width $\sigma_{w}$ of the
spindle-shaped function $\frac{\delta P_{2b}}{\delta\chi(x,z)}$,
and we also used the Poisson distribution formula to calculate the
RMS error of the counting measurement of $\frac{\delta P_{2b}}{\delta\chi(x,z)}$
as $\frac{1}{\sqrt{n}}$ which is to be used to check against the
experiment. In section 4.6, the estimate shows the experiment is feasible.

We proposed and analyzed the basic principle of an experiment using
photons at a wavelength of $0.5\mu m$. Many aspects of the problem
need to be studied further. For example, it would be interesting to
study the feasibility of other wavelengths or particles. It would
also be interesting to study the effect of the photon's finite bandwidth
and pulse length and see if it will affect the width of the spindle-shaped
region. The experiment can produce high-resolution images only using
low-intensity beams, and we expect it might find applications.
\begin{acknowledgments}
We thank Professor C. N. Yang for drawing our attention to the problem
of dissipative systems, for spending his valuable time in many sessions
of stimulating discussions on this subject, and for many suggestions
that are critically important for work on the dissipative systems
\cite{yu}, which finally leads to the topics in the present work.
We thank Dr. T. Shaftan, and Dr. V. Smaluk for their discussion and
suggestions on the manuscript. We thank Prof. Shih for the discussion
and suggestions on the manuscript.
\end{acknowledgments}

\section*{Appendix I Derive Green's function $G(x_{1},x_{0};t)$ in Heinsenberg
Picture }

We derive Green's function $G(x_{1},x_{0};t)$ in the Heisenberg Picture
using the method used in our work on dissipative systems \cite{yu}.�Instead
of using $x_{1},t_{1},x_{2},t_{2}$ as position and time at slits
1 and 2 as in Fig. 1, we use initial position and time $x_{0},0$
and $x,t$ to avoid confusing the derivation with the derivation elsewhere,
so that when we apply the Green's function, we replace the indices.
We write

\begin{equation}
x(t)=x_{0}+\frac{p_{0}}{M}t=x_{0}-t\frac{i\hbar}{M}\frac{\partial}{\partial x_{0}}\label{xteq}
\end{equation}

The eigenfunction of $x(t)$ with an eigenvalue denoted by $x_{1}$,
in the $x_{0}$ representation, is easily calculated to be

\begin{equation}
u_{x_{1}}(x_{0},t)=\left(\frac{M}{2\pi\hbar t}\right)^{\frac{1}{2}}\exp\left[-i\frac{M}{2\hbar t}\left(x_{0}^{2}-2x_{1}x_{0}+\phi(x_{1},t)\right)\right],\label{ux1}
\end{equation}

with $\phi$ as an arbitrary phase, i.e., a real number. This eigenfunction
is related to Green's function $G(x_{1},x_{0};t)=$ $<x_{1}|U(t)|x_{0}>$
where $U(t)$ is the evolution operator. To see this, we use the relation
between the Schoedinger operator $\mathbf{X}_{S}\equiv x_{0}$ and
the Heisenberg operator $x(t)\equiv\mathbf{X}_{H}(t)=U^{-1}(t)\mathbf{X}_{S}U(t)$. 

Let $|x_{1}>$ be the eigenvector of $\mathbf{X}_{S}$ with eigenvalue
$x_{1}$, i.e.,

\begin{align}
 & \mathbf{X}_{S}|x_{1}>=x_{1}|x_{1}>\label{eigenvector}
\end{align}

we see that $U^{-1}|x_{1}>$ is the eigenvector of $x(t)$ of value
$x_{1}$

\begin{align}
 & x(t)U^{-1}|x_{1}>=\mathbf{X}_{H}U^{-1}|x_{1}>=U^{-1}\mathbf{X}_{S}UU^{-1}|x_{1}>=U^{-1}\mathbf{X}_{S}|x_{1}>=x_{1}U^{-1}|x_{1}>\label{eigenvectorofXt}
\end{align}

So $U^{-1}|x_{1}>$ is the eigenvector of $X(t)$, thus it is proportional
to $u_{x_{1}}(x_{0},t)$. We have $u_{x1}(x_{0},t)=<x_{0}|U^{-1}(t)|x_{1}>$
$=<x_{0}|U^{\dagger}(t)|x_{1}>=G^{*}(x_{1},x_{0};t)$, where the evolution
operator by $U(t)$ is unitary when we choose the eigenvectors of
$x(t)$ to be orthonormal. Thus we have

\begin{equation}
G(x_{1},x_{0};t)=u_{x_{1}}^{*}(x_{0},t)=\left(\frac{M}{2\pi\hbar t}\right)^{\frac{1}{2}}\exp\left[i\frac{M}{2\hbar t}\left(x_{0}^{2}-2x_{1}x_{0}+\phi(x_{1},t)\right)\right]\label{greens_function}
\end{equation}

Next, we shall determine the arbitrary phase $\phi(x_{1},t)$ , the
phase of the eigenvectors of $x(t)$ in Eq. (\ref{ux1}). Since $p$
is constant in free space, $p(t)=p_{0}$

The eigenfunction of $p(t)$ can be calculated in two ways. (a) We
can calculate the eigenvector of $p(t)=p_{0}=-\frac{i\hbar}{M}\frac{\partial}{\partial x_{0}}$
in the $x_{0}$ representation and then use the Green's function Eq.
(\ref{greens_function}) to transform it into the $x(t)$ representation.
(b) The eigenfunction of $p(t)$ with eigenvalue $p_{1}$ is $e^{i\frac{p_{1}}{\hbar}x_{1}}$.
By comparing these two solutions, the arbitrary phase $\phi(x_{1},t)$
in the Green's function is determined to be within a phase $\phi(t)$,
which is independent of $x_{1}$. $\phi(t)$ is an arbitrary real
function of time, except that $\phi(0)=0$ so that it satisfies the
condition that at $t=0$, the Green's function becomes $\delta(x_{1}-x_{0})$. 

\begin{equation}
G(x_{1},x_{0};t)=\left(\frac{M}{2\pi i\hbar t}\right)^{\frac{1}{2}}\exp\left[i\frac{M}{2\hbar t}\left(x_{0}^{2}-2x_{1}x_{0}+x_{1}^{2}\right)-\frac{i}{\hbar}\phi(t)\right]
\end{equation}

When substituting $G(x_{1},x_{0};t)$ into the Schr\textcyr{\"\cyro}dinger
equation (\ref{1DschoedingerEq}), it requires $\dot{\phi}(t)=0$.
Since $\phi(0)=0$, $\phi(t)=0$. Thus we obtain the Green's function:

\begin{equation}
G(x_{1},x_{0};t)=\left(\frac{M}{2\pi i\hbar t}\right)^{\frac{1}{2}}\exp\left[i\frac{M}{2\hbar t}\left(x_{1}-x_{0}\right)^{2}\right]
\end{equation}

\section*{Appendix II Wave Function Propagation From Slit 1 to Slit 2}

\paragraph*{II.1 Derive $\ln\psi_{2a}(x_{2},t_{1})$ and $\ln\psi_{2b}(x_{2},t_{1})$ }

We apply the Green's function Eq.(\ref{greens_function-1}) in the
Eq.(\ref{propagation})

\begin{align}
 & \psi_{2a}(x_{2},z_{2})=\int dx_{1}G(x_{2},x_{1};t_{2}-t_{1})\psi_{1}(x_{1},z_{1})\label{propagation-1}\\
 & \psi_{2b}(x_{2},z_{2})=f_{2}(x_{2})\psi_{2a}(x_{2},z_{2})\nonumber 
\end{align}

to calculate the wave function $\ln\psi_{2a}(x_{2},t_{1})$ and $\ln\psi_{2b}(x_{2},t_{1})$.

For the first step of propagation, because both $\ln\psi_{1}(x_{1},z_{1})$
and $\ln G(x_{2},x_{1};t_{2}-t_{1})$ are quadratic forms of $x_{1}$,
this is a Gaussian integral. Use the notations of section 3.1: $z_{j}\equiv vt_{j},$$\frac{Mv}{\hbar}=k=\frac{2\pi}{\lambda}$
and $z=z_{2}-z_{1}=v(t_{2}-t_{1})$ in Fig. 1, and Eq.(\ref{parameters}),
we have 

\begin{align}
 & \ln\psi_{1}(x_{1},z_{1}=0)=\alpha_{1}(x_{1}-s_{1})^{2}+c_{1}\nonumber \\
 & \alpha_{1}=-\frac{1}{4\sigma_{1}^{2}},c_{1}=\frac{1}{4}\ln\left(\frac{1}{2\pi\sigma_{1}^{2}}\right)\nonumber \\
 & \ln G(x_{2},x_{1};t_{2}-t_{1})=i\beta_{1}(x_{2}-x_{1})^{2}+\frac{1}{2}\ln\frac{\beta_{1}}{i\pi}\label{green's function}\\
 & \beta_{1}=\frac{M}{2\hbar(t_{2}-t_{1})}\equiv\frac{1}{4d_{1}},d_{1}=\frac{\hbar(t_{2}-t_{1})}{2M}=\frac{z_{2}-z_{1}}{2k}=\frac{\lambda L}{4\pi}\nonumber \\
 & \psi_{2a}(x_{2},z_{2})=\int dx_{1}\exp(\alpha_{1}(x_{1}-s_{1})^{2}+i\beta_{1}(x_{2}-x_{1})^{2}+c_{1}+\frac{1}{2}\ln\frac{\beta_{1}}{i\pi})\nonumber 
\end{align}

The exponent of the integrand in $\psi_{2a}(x_{2},z_{2})$ can be
written into a standard quadratic form of $ax_{1}^{2}+bx_{1}+c$.
It is then integrated out using the following Gaussian integral, which
is to be applied repeatedly in the next steps, as long as the real
part of $a$ is negative.

\begin{equation}
\int dx\exp\left(ax^{2}+bx+c\right)=\left(\frac{\pi}{-a}\right)^{\frac{1}{2}}\exp\left(-\frac{b^{2}}{4a}+c\right)=\exp\left(-\frac{b^{2}}{4a}+c+\ln\left(\frac{\pi}{-a}\right)^{\frac{1}{2}}\right),\label{gaussian_integral}
\end{equation}

In the following applications of this formula, we always checked that
indeed $\text{Re}(a)<0$. The result is written into the same standard
quadratic form as $\ln\psi_{1}(x_{1},z_{1})$, but in terms of the
variable $x_{2}$ 
\begin{align}
 & \ln\psi_{2a}(x_{2},t)=\alpha_{2a}\left(x_{2}-s_{2a}\right)^{2}+c_{2a}\label{propagation formula}\\
 & \alpha_{2a}=\frac{i\alpha_{1}\beta_{1}}{\left(\alpha_{1}+i\beta_{1}\right)}=\frac{1}{\frac{1}{\alpha_{1}}+\frac{1}{i\beta_{1}}}=-\frac{1}{4}\frac{1}{\sigma_{1}^{2}+id_{1}}\equiv-\frac{1}{4\Delta_{1}}\nonumber \\
 & s_{2a}=s_{1}\nonumber \\
 & c_{2a}=\frac{1}{2}\ln\left(\frac{\alpha_{2a}}{\alpha_{1}}\right)+c_{1}=\frac{1}{2}\ln\left(\frac{\sigma_{1}^{2}}{\sigma_{1}^{2}+id_{1}}\right)+\frac{1}{4}\ln\left(\frac{1}{2\pi\sigma_{1}^{2}}\right)\nonumber 
\end{align}

Here we defined a parameter in Eq.(\ref{parameters}), $\Delta_{1}\equiv\sigma_{1}^{2}+id_{1}$
so that $\alpha_{2a}=-\frac{1}{4\Delta_{1}}$ is similar to $\alpha_{1}=-\frac{1}{4\sigma_{1}^{2}}$,
with the width $\sigma_{1}^{2}$ replaced by $\Delta_{1}$. 

The next step is to pass through the slit 2 with $f_{2}(x_{2})=\exp(\alpha_{f2}(q_{2}-s_{2})^{2})$,
$\alpha_{f2}\equiv-\frac{1}{4\sigma_{2}^{2}}$ , the result of applying
$\psi(x_{2})=f_{2}(x_{2})\psi_{2a}(x_{2})$ is

\begin{align}
 & \ln\psi_{2b}(x_{2},z_{2}=z)=\alpha_{2b}(x_{2}-s_{2b})^{2}+c_{2b}\label{slit_formula}\\
 & \alpha_{2b}=\alpha_{2a}+\alpha_{f2}\nonumber \\
 & s_{2b}=\frac{\alpha_{2a}s_{2a}+\alpha_{f2}s_{2}}{\alpha_{2a}+\alpha_{f2}}\nonumber \\
 & c_{2b}=-\alpha_{2b}s_{2b}^{2}+\alpha_{2a}s_{2a}^{2}+\alpha_{f2}s_{2}^{2}+c_{2a}\nonumber 
\end{align}

The next steps are to derive more explicit expressions for $\ln\psi_{2a}(x_{2},t)$
and $\ln\psi_{2b}(x_{2},z_{2}=z)$.

\paragraph*{II.2 $\alpha$ formula, }

Apply the two formulas of II.1, we find,

\begin{align}
 & \frac{1}{\alpha_{2a}}=\frac{1}{\alpha_{1}}-i\frac{1}{\beta_{1}}\label{alpha pattern-1}\\
 & \alpha_{2b}=\alpha_{2a}+\alpha_{f2}\nonumber 
\end{align}

Applying Eq.(\ref{green's function}), Eq.(\ref{propagation formula}),
and Eq.(\ref{slit_formula}) in Eq.(\ref{alpha pattern-1}), we have

\begin{align}
 & \frac{1}{\alpha_{1}}=-4\sigma_{1}^{2}\nonumber \\
 & \frac{1}{\alpha_{2a}}=-4\left(\sigma_{1}^{2}+id_{1}\right)\equiv-4\Delta_{1},\alpha_{2b}=\alpha_{2a}+\alpha_{f2}=-\frac{1}{4\Delta_{1}}-\frac{1}{4\sigma_{2}^{2}}\\
 & \frac{1}{\alpha_{2b}}=-\frac{1}{\frac{1}{4\Delta_{1}}+\frac{1}{4\sigma_{2}^{2}}}=-4\sigma_{2}^{2}\frac{\Delta_{1}}{\sigma_{2}^{2}+\Delta_{1}}\label{alpha_pattern-1}
\end{align}

\paragraph*{II.3 $s$-formula}

Follow Eq.(\ref{parameters},\ref{psi_pattern}), and use $\alpha$
formula of I.2, we have, 

\begin{align}
 & s_{2a}=s_{1}\nonumber \\
 & s_{2b}=\frac{\alpha_{2a}s_{2a}+\alpha_{f2}s_{2}}{\alpha_{2a}+\alpha_{f2}}=\frac{\left(\frac{s_{2a}}{\alpha_{f2}}+\frac{s_{2}}{\alpha_{2a}}\right)}{\left(\frac{1}{\alpha_{2a}}+\frac{1}{\alpha_{f2}}\right)}=\frac{-4\sigma_{2}^{2}s_{2a}-4\Delta_{1}s_{2}}{\left(-4\Delta_{1}-4\sigma_{2}^{2}\right)}=\frac{\sigma_{2}^{2}s_{2a}+\Delta_{1}s_{2}}{\Delta_{1}+\sigma_{2}^{2}}\nonumber \\
 & =\frac{\sigma_{2}^{2}s_{1}+\Delta_{1}s_{2}}{\Delta_{1}+\sigma_{2}^{2}}=s_{2}\frac{\Delta_{1}}{\Delta_{1}+\sigma_{2}^{2}}+\frac{\Delta_{1}s_{1}+\sigma_{2}^{2}s_{1}-\Delta_{1}s_{1}}{\Delta_{1}+\sigma_{2}^{2}}\label{s_steps}\\
 & =\left(s_{2}-s_{1}\right)\frac{\Delta_{1}}{\Delta_{1}+\sigma_{2}^{2}}+s_{1}\nonumber 
\end{align}

\paragraph*{II.4 $c$ formula}

Using Eq.(\ref{alpha_pattern-1}) and Eq.(\ref{slit_formula},\ref{s_steps}),
we use the following steps to find explicit expressions for $c_{2b}$,

\begin{align}
 & c_{2b}=-\alpha_{2b}s_{2b}^{2}+\alpha_{2a}s_{2a}^{2}+\alpha_{f2}s_{2}^{2}+c_{2a}\nonumber \\
 & \alpha_{2a}=-\frac{1}{4\Delta_{1}},\alpha_{2b}=-\frac{1}{4\Delta_{1}}\frac{\Delta_{1}+\sigma_{2}^{2}}{\sigma_{2}^{2}},\alpha_{f2}=-\frac{1}{4\sigma_{2}^{2}}\nonumber \\
 & c_{2b}=\frac{1}{4\Delta_{1}}\frac{\Delta_{1}+\sigma_{2}^{2}}{\sigma_{1}^{2}}s_{2b}^{2}-\frac{1}{4\Delta_{1}}s_{2a}^{2}-\frac{1}{4\sigma_{2}^{2}}s_{2}^{2}+c_{2a}\label{c_steps}\\
 & s_{2b}=\frac{\sigma_{2}^{2}s_{2a}+\Delta_{1}s_{2}}{\Delta_{1}+\sigma_{2}^{2}},s_{2a}=s_{1}\nonumber \\
 & c_{2b}{\displaystyle =-\frac{s_{2}^{2}}{4\sigma_{2}^{2}}-\frac{s_{1}^{2}}{4\Delta_{1}}+\frac{\left(\Delta_{1}s_{2}+s_{1}\sigma_{2}^{2}\right)^{2}}{4\Delta_{1}\sigma_{2}^{2}\left(\Delta_{1}+\sigma_{2}^{2}\right)}}+c_{2a}={\displaystyle -\frac{\left(s_{1}-s_{2}\right)^{2}}{4\Delta_{1}+4\sigma_{2}^{2}}}+c_{2a}\nonumber 
\end{align}

Thus we have $c$ formula,

\begin{align*}
 & c_{1}=\frac{1}{4}\ln\left(\frac{1}{2\pi\sigma_{1}^{2}}\right),c_{2a}=\frac{1}{2}\ln\left(\frac{\alpha_{2a}}{\alpha_{1}}\right)+c_{1}=\frac{1}{2}\ln\left(\frac{\sigma_{1}^{2}}{\Delta_{1}}\right)+\frac{1}{4}\ln\left(\frac{1}{2\pi\sigma_{1}^{2}}\right)\\
 & c_{2b}=-\frac{(s_{1}-s_{2})^{2}}{4\left(\Delta_{1}+\sigma_{2}^{2}\right)}+c_{2a}=-\frac{(s_{1}-s_{2})^{2}}{4\left(\Delta_{1}+\sigma_{2}^{2}\right)}+\frac{1}{2}\ln\left(\frac{\sigma_{1}^{2}}{\Delta_{1}}\right)+\frac{1}{4}\ln\left(\frac{1}{2\pi\sigma_{1}^{2}}\right)
\end{align*}

\paragraph*{II.5 $\psi$ formula}

All the steps above applied to Eq.(\ref{propagation}) lead to the
$\psi$ formula,

\begin{align}
 & \ln\psi_{1}(x_{1},0)=\alpha_{1}(x_{1}-s_{1})^{2}+c_{1}=-\frac{1}{4\sigma_{1}^{2}}(x_{1}-s_{1})^{2}+\frac{1}{4}\ln\left(\frac{1}{2\pi\sigma_{1}^{2}}\right)\nonumber \\
 & \ln\psi_{2a}(x_{2},t_{1})=\alpha_{2a}\left(x_{2}-s_{2a}\right)^{2}+c_{2a}=-\frac{1}{4\Delta_{1}}\left(x_{2}-s_{1}\right)^{2}+\frac{1}{2}\ln\left(\frac{\sigma_{1}^{2}}{\Delta_{1}}\right)+\frac{1}{4}\ln\left(\frac{1}{2\pi\sigma_{1}^{2}}\right)\nonumber \\
 & \ln\psi_{2b}(x_{2},t_{1})=\alpha_{2b}\left(x_{2}-s_{2b}\right)^{2}+c_{2b}=-\frac{1}{4\Delta_{1}}\frac{\Delta_{1}+\sigma_{2}^{2}}{\sigma_{2}^{2}}\left(x_{2}-s_{2b}\right)^{2}-\frac{(s_{1}-s_{2})^{2}}{4\left(\Delta_{1}+\sigma_{2}^{2}\right)}+\frac{1}{2}\ln\left(\frac{\sigma_{1}^{2}}{\Delta_{1}}\right)+\frac{1}{4}\ln\left(\frac{1}{2\pi\sigma_{1}^{2}}\right)\label{psi-pattern}
\end{align}

where,

\begin{align*}
 & s_{2b}=\left(s_{2}-s_{1}\right)\frac{\Delta_{1}}{\Delta_{1}+\sigma_{2}^{2}}+s_{1}\\
 & \Delta_{1}\equiv\sigma_{1}^{2}+id_{1},\beta_{1}=\frac{1}{4d_{1}},d_{1}=\frac{\hbar(t_{2}-t_{1})}{2M}=\frac{z_{2}-z_{1}}{2k}=\frac{\lambda L}{4\pi}
\end{align*}

\section*{Appendix III simplified analytical expression of $P_{2b}$ }

The expression of $\ln\psi_{2b}(x_{2},z_{2})$ in Eq.(\ref{psi_pattern})
is again a quadratic form of $x_{2}$, hence $\ln\psi_{2b}(x_{2},z_{2})+\ln\psi_{2b}^{*}(x_{2},z_{2})$
in the integrand of $P_{2b}=\int dq_{2}|\psi_{2b}(x_{2})|^{2}$ is
also a quadratic form. Thus Eq.(\ref{gaussian_integral}) can be applied
as a Gaussian integral to get,

\begin{align}
 & P_{2b}(s_{2},\sigma_{2},z_{2})=\sqrt{\frac{\pi}{-2\text{Re}(\alpha_{2b})}}\exp\left(-2{\displaystyle \frac{|\alpha_{2b}|^{2}\left(\text{Im(}s_{2b})\right)^{2}}{\text{Re}(\alpha_{2b})}}+2\text{Re}\left(c_{2b}\right)\right)\label{P2bAII}
\end{align}

This $P_{2b}$ expression is analytical but expressed by $\alpha_{2b},s_{2b},c_{2b}$
that in turn are represented by $\alpha_{2a},s_{2a},c_{2a}$, not
by the parameters directly used in the experiment,

When we express it using these direct parameters by a series of substitutions,
it becomes a long and complicated expression that needs to be simplified.
We shall outline some intermediate steps to avoid writing down tedious
derivation and cluttering. First, we substitute $c_{2b}$ from Eq.(\ref{psi_pattern}),
and simplify the real and imaginary parts of the results in terms
of two scaled dimensionless parameters, $\mu,\rho$

\begin{align}
 & P_{2b}=\sqrt{\frac{\pi}{-2\text{Re}(\alpha_{2b})}}\exp\left({\displaystyle \frac{|\alpha_{2b}|^{2}\left(s_{2b}-s_{2b}^{*}\right)^{2}}{2\text{Re}(\alpha_{2b})}}+2\text{Re}\left(-\frac{(s_{1}-s_{2})^{2}}{4\left(\Delta_{1}+\sigma_{2}^{2}\right)}+\frac{1}{2}\ln\left(\frac{\sigma_{1}^{2}}{\Delta_{1}}\right)+\frac{1}{4}\ln\left(\frac{1}{2\pi\sigma_{1}^{2}}\right)\right)\right)\nonumber \\
 & \mu\equiv\frac{\sigma_{1}^{2}}{d_{1}},\rho\equiv\frac{\sigma_{2}^{2}}{\sigma_{1}^{2}}\nonumber \\
 & \text{Re}(\alpha_{2b})=\frac{1}{2}\left(-\frac{1}{4\sigma_{2}^{2}}\frac{\Delta_{1}+\sigma_{2}^{2}}{\Delta_{1}}-\frac{1}{4\sigma_{2}^{2}}\frac{\Delta_{1}^{*}+\sigma_{2}^{2}}{\Delta_{1}^{*}}\right)=-\frac{1}{4\sigma_{2}^{2}}\frac{\sigma_{1}^{4}+d_{1}^{2}+\sigma_{1}^{2}\sigma_{2}^{2}}{\sigma_{1}^{4}+d_{1}^{2}}=-\frac{1}{4\sigma_{2}^{2}}\frac{\mu^{2}+\rho\mu^{2}+1}{\mu^{2}+1}\label{realpha2b}\\
 & \exp\left(2\text{Re}\left(\frac{1}{2}\ln\left(\frac{\sigma_{1}^{2}}{\Delta_{1}}\right)\right)\right)=\exp\left(\frac{1}{2}\ln\left(\frac{\sigma_{1}^{2}}{\Delta_{1}}\right)+\frac{1}{2}\ln\left(\frac{\sigma_{1}^{2}}{\Delta_{1}^{*}}\right)\right)=\sqrt{\frac{\sigma_{1}^{4}}{\sigma_{1}^{4}+d_{1}^{2}}}=\sqrt{\frac{\mu^{2}}{\mu^{2}+1}}\nonumber \\
 & \therefore P_{2b}=\sqrt{\frac{\rho\mu^{2}}{{\displaystyle \mu^{2}+\mu^{2}\rho+1}}}\exp\left({\displaystyle \frac{|\alpha_{2b}|^{2}\left(s_{2b}-s_{2b}^{*}\right)^{2}}{2\text{Re}(\alpha_{2b})}}+2\text{Re}\left(-\frac{(s_{1}-s_{2})^{2}}{4\left(\Delta_{1}+\sigma_{2}^{2}\right)}\right)\right)\nonumber 
\end{align}

We need to simplify $|\alpha_{2b}|^{2}\left(s_{2b}-s_{2b}^{*}\right)^{2}+4\text{Re}\left(-\frac{(s_{1}-s_{2})^{2}}{4\left(\Delta_{1}+\sigma_{2}^{2}\right)}\right)\text{Re}(\alpha_{2b})$,
and we have

\begin{align}
 & |\alpha_{2b}|^{2}\left(s_{2b}-s_{2b}^{*}\right)^{2}=\frac{1}{16}\frac{1}{\Delta_{1}\Delta_{1}^{*}}\frac{\left(\Delta_{1}-\Delta_{1}^{*}\right)^{2}}{\left(\Delta_{1}+\sigma_{2}^{2}\right)\left(\Delta_{1}^{*}+\sigma_{2}^{2}\right)}\left(s_{2}-s_{1}\right)^{2}\\
 & 4\text{Re}\left(-\frac{(s_{1}-s_{2})^{2}}{4\left(\Delta_{1}+\sigma_{2}^{2}\right)}\right)\text{Re}(\alpha_{2b})=\frac{1}{16\sigma_{2}^{2}}(s_{1}-s_{2})^{2}\left(\frac{\Delta_{1}+\sigma_{2}^{2}+\Delta_{1}^{*}+\sigma_{2}^{2}}{\left(\Delta_{1}+\sigma_{2}^{2}\right)\left(\Delta_{1}^{*}+\sigma_{2}^{2}\right)}\right)\left(\frac{2\Delta_{1}\Delta_{1}^{*}+\sigma_{2}^{2}(\Delta_{1}+\Delta_{1}^{*})}{\Delta_{1}\Delta_{1}^{*}}\right)\nonumber 
\end{align}

their sum can be simplified, and using $\text{Re}(\alpha_{2b})$ in
Eq.(\ref{realpha2b}), we have

\begin{align}
 & |\alpha_{2b}|^{2}\left(s_{2b}-s_{2b}^{*}\right)^{2}+4\text{Re}\left(-\frac{(s_{1}-s_{2})^{2}}{4\left(\Delta_{1}+\sigma_{2}^{2}\right)}\right)\text{Re}(\alpha_{2b})=\frac{1}{8\sigma_{2}^{2}}(s_{1}-s_{2})^{2}\frac{\left(\Delta_{1}+\Delta_{1}^{*}\right)}{\Delta_{1}\Delta_{1}^{*}}=\frac{\sigma_{1}^{2}}{4\sigma_{1}^{4}\sigma_{2}^{2}}\frac{\sigma_{1}^{4}}{\sigma_{1}^{4}+d_{1}^{2}}(s_{1}-s_{2})^{2}\\
 & {\displaystyle \frac{|\alpha_{2b}|^{2}\left(s_{2b}-s_{2b}^{*}\right)^{2}}{2\text{Re}(\alpha_{2b})}}+2\text{Re}\left(-\frac{(s_{1}-s_{2})^{2}}{4\left(\Delta_{1}+\sigma_{2}^{2}\right)}\right)={\displaystyle \frac{1}{2\text{Re}(\alpha_{2b})}}\frac{\sigma_{1}^{2}}{4\sigma_{1}^{4}\sigma_{2}^{2}}\frac{\sigma_{1}^{4}}{\sigma_{1}^{4}+d_{1}^{2}}(s_{1}-s_{2})^{2}\nonumber \\
 & =-\frac{1}{2}\frac{4\sigma_{2}^{2}\left(\sigma_{1}^{4}+d_{1}^{2}\right)}{\sigma_{1}^{4}+d_{1}^{2}+\sigma_{1}^{2}\sigma_{2}^{2}}\frac{\sigma_{1}^{2}}{4\sigma_{1}^{4}\sigma_{2}^{2}}\frac{\sigma_{1}^{4}}{\sigma_{1}^{4}+d_{1}^{2}}(s_{1}-s_{2})^{2}=-\frac{1}{2}\frac{\sigma_{1}^{2}}{\sigma_{1}^{4}+d_{1}^{2}+\sigma_{1}^{2}\sigma_{2}^{2}}(s_{1}-s_{2})^{2}=-\frac{1}{2\sigma_{1}^{2}}\frac{\mu^{2}}{\mu^{2}+1+\rho\mu^{2}}(s_{1}-s_{2})^{2}\nonumber \\
 & \therefore P_{2b}=\sqrt{\frac{\rho\mu^{2}}{{\displaystyle \mu^{2}+\rho\mu^{2}+1}}}\exp\left(-\frac{1}{2\sigma_{1}^{2}}\frac{\mu^{2}}{\mu^{2}+1+\rho\mu^{2}}(s_{1}-s_{2})^{2}\right)\nonumber 
\end{align}

In the limit of small slit sizes, as $\sigma_{1},\sigma_{2}\rightarrow0$,
we need to specify their ratio $\rho=\frac{\sigma_{2}^{2}}{\sigma_{1}^{2}}$,
we find 

\begin{equation}
P_{2b}=\sqrt{\rho\mu^{2}}\exp\left(-\frac{\mu^{2}}{2\sigma_{1}^{2}}(s_{1}-s_{2})^{2}\right)=\frac{\sigma_{1}\sigma_{2}}{d_{1}}\exp\left(-\frac{\sigma_{1}^{2}}{2d_{1}^{2}}(s_{1}-s_{2})^{2}\right)
\end{equation}

\section*{Appendix IV analytical expression of $\frac{\delta P_{2b}}{\delta\chi(x,z)}=\phi_{b}(x,z)\psi(x,z)+c.c.$ }

\subsection*{IV.1 Derivation of the perturbative function $\frac{\delta}{\delta\chi(x,z)}P_{2b}$ }

To calculate $\frac{\delta}{\delta\chi(x,z)}P_{2b}$ using Eq.(\ref{perturbative_function}), 

\begin{align}
 & \frac{\delta P_{2b}}{\delta\chi(x,z)}=\phi_{b}(x)\psi(x)+c.c.\label{perturbative_function-1}\\
 & \psi(x)\equiv\int_{-\infty}^{\infty}dx_{1}G(x,x_{1};t)\psi_{1}(x_{1})\nonumber \\
 & \phi_{b}(x)\equiv\int_{-\infty}^{\infty}dx_{2}\psi_{2b}^{*}(x_{2},z_{2})f_{2}(x_{2})G(x_{2},x;t_{2}-t)\nonumber 
\end{align}
we first need to express $\psi_{2b}(x)$ and $\psi(x)$ in terms of
the parameters of the experimental setup more directly. We have $\ln\psi_{2b}(x_{2},z_{2})$
in Eq.(\ref{psi_pattern}), 

\begin{align}
 & \ln\psi_{2b}(x_{2},z_{2}=L)=\alpha_{2b}\left(x_{2}-s_{2b}\right)^{2}+c_{2b}=-\frac{1}{4\Delta_{1}}\frac{\Delta_{1}+\sigma_{2}^{2}}{\sigma_{2}^{2}}\left(x_{2}-s_{2b}\right)^{2}-\frac{(s_{1}-s_{2})^{2}}{4\left(\Delta_{1}+\sigma_{2}^{2}\right)}+\frac{1}{2}\ln\left(\frac{\sigma_{1}^{2}}{\Delta_{1}}\right)+\frac{1}{4}\ln\left(\frac{1}{2\pi\sigma_{1}^{2}}\right)\label{psi2b}\\
 & \Delta_{1}=\sigma_{1}^{2}+id_{1},d_{1}=\frac{\lambda(z_{2}-z_{1})}{4\pi}=\frac{\lambda L}{4\pi}\nonumber 
\end{align}

Because $f_{2}(x_{2}),\alpha_{f2}$ are the same as given in section
3.2, i.e., $f_{2}(x_{2})=\exp(\alpha_{f2}(q_{2}-s_{2})^{2}),\alpha_{f2}\equiv-\frac{1}{4\sigma_{2}^{2}}$,
defining $\beta_{2}=\frac{1}{4d_{2}},d_{2}=\frac{\lambda(L-z)}{4\pi}$,
and applying Eq.(\ref{greens_function-1}) for the Green's function,
we have 

\begin{align}
 & \ln f_{2}(x_{2})G(x_{2},x;t_{2}-t)=\alpha_{f2}(x_{2}-s_{2})^{2}+i\beta_{2}(x_{2}-x)^{2}+\frac{1}{2}\ln\frac{\beta_{2}}{i\pi}\nonumber \\
 & \ln\psi_{2b}^{*}(x_{2},z_{2})f_{2}(x_{2})G(x_{2},x;t_{L}-t)=\alpha_{f2}(x_{2}-s_{2})^{2}+i\beta_{2}(x_{2}-x)^{2}+\overline{\alpha_{2b}}\left(x_{2}-\overline{s_{2b}}\right)^{2}+\frac{1}{2}\ln\frac{\beta_{2}}{i\pi}+\overline{c_{2b}}\label{lnpsi2bs}\\
 & =\left(\alpha_{f2}+i\beta_{2}+\overline{\alpha_{2b}}\right)x_{2}^{2}-2\left(\alpha_{f2}s_{2}+i\beta_{2}x+\overline{\alpha_{2b}}\overline{s_{2b}}\right)x_{2}+\alpha_{f2}s_{2}^{2}+i\beta_{2}x^{2}+\overline{\alpha_{2b}}\overline{s_{2b}}^{2}+\frac{1}{2}\ln\frac{\beta_{2}}{i\pi}+\overline{c_{2b}}\nonumber 
\end{align}

Apply the Gaussian integral $\int dx\exp\left(ax^{2}+bx+c\right)=\exp\left(-\frac{b^{2}}{4a}+c+\ln\left(\frac{\pi}{-a}\right)^{\frac{1}{2}}\right)$,

\begin{align}
 & \ln\phi_{b}(x)=\ln\int dx_{2}\psi_{2b}^{*}(x_{2},z_{2})f_{2}(x_{2})G(x_{2},x;t_{2}-t)\nonumber \\
 & =\alpha_{f2}s_{2}^{2}+i\beta_{2}x^{2}+\frac{1}{2}\ln\frac{\beta_{2}}{i\pi}+\ln\left(\frac{\pi}{-\left(\alpha_{f2}+i\beta_{2}+\overline{\alpha_{2b}}\right)}\right)^{\frac{1}{2}}+\overline{\alpha_{2b}}\overline{s_{2b}}^{2}+\overline{c_{2b}}-\frac{4\left(\alpha_{f2}s_{2}+i\beta_{2}x+\overline{\alpha_{2b}}\overline{s_{2b}}\right)^{2}}{4\left(\alpha_{f2}+i\beta_{2}+\overline{\alpha_{2b}}\right)}\label{lnphib}
\end{align}

and because in $\psi(x)\equiv\int dx_{1}G(x,x_{1};t)\psi_{1}(x_{1})$,
the propagation distance from the entrance slit to the pin is $z$,
not $z_{2}$, from Eq.(\ref{psi_pattern},\ref{parameters}) we have 

\begin{align}
 & {\displaystyle \ln\psi(x)=\frac{\ln\left(\frac{1}{2\pi\sigma_{1}^{2}}\right)}{4}+\frac{\ln\left(\frac{\sigma_{1}^{2}}{\Delta}\right)}{2}-\frac{\left(x-s_{1}\right)^{2}}{4\Delta}}\label{lnphi2a}\\
 & \Delta=\sigma_{1}^{2}+id,d=\frac{\lambda z}{4\pi},\nonumber 
\end{align}

Since $\ln\phi_{b}(x)$ and $\ln\psi(x)$ are both quadratic forms
of $x$, $\phi_{b}(x)\psi(x)$ is Gaussian. we can write $\ln\phi_{b}(x)\psi(x)$
in the quadratic form as 

\begin{align}
 & \ln\phi_{b}(x)\psi(x)=\alpha_{\chi}\left(x-x_{c}\right)^{2}+C\nonumber \\
 & \alpha_{\chi}=i\beta_{2}-A\left(i\beta_{2}\right)^{2}-\frac{1}{4\Delta}\nonumber \\
 & x_{c}=\frac{\left(A_{2}i\beta_{2}\left(\alpha_{f2}s_{2}+\overline{\alpha_{2b}}\overline{s_{2b}}\right)-s_{1}\frac{1}{4\Delta}\right)}{\alpha_{\chi}}\label{lnphipsi}\\
 & C=-\frac{\left(A_{2}i\beta_{2}\left(\alpha_{f2}s_{2}+\overline{\alpha_{2b}}\overline{s_{2b}}\right)-s_{1}\frac{1}{4\Delta}\right)^{2}}{\alpha_{\chi}}-A_{2}\left(\alpha_{f2}s_{2}+\overline{\alpha_{2b}}\overline{s_{2b}}\right)^{2}-\frac{\left(s_{1}\right)^{2}}{4\Delta}+\alpha_{f2}s_{2}^{2}\nonumber \\
 & +\frac{1}{2}\ln\frac{\beta_{2}}{i\pi}+\ln\left(\left(-\pi A_{2}\right)^{\frac{1}{2}}\right)+\overline{\alpha_{2b}}\overline{s_{2b}}^{2}+\overline{c_{2b}}+\frac{\ln\left(\frac{1}{2\pi\sigma_{1}^{2}}\right)}{4}+\frac{\ln\left(\frac{\sigma_{1}^{2}}{\Delta}\right)}{2}\nonumber \\
 & A\equiv\frac{1}{\alpha_{f2}+i\beta_{2}+\overline{\alpha_{2b}}}\nonumber 
\end{align}

where we introduce $A$ to simplify the writing. Thus we have

\begin{align}
 & \frac{\delta P_{2b}}{\delta\chi(x)}=\phi_{b}(x)\psi(x)+c.c.\label{phipsi}\\
 & =\exp(\alpha_{\chi}\left(x-x_{c}\right)^{2}+C)+c.c.\nonumber 
\end{align}

Where $\alpha_{\chi}$ and $x_{c}$ are identified by completing the
quadratic form and simplified. To make writing simpler, we define
the dimensionless parameter of distance ratio $\xi$ as follows: we
find (derivation of $\alpha_{\chi}$ is given in Appendix IV.2),

\begin{align}
 & \alpha_{\chi}=-\frac{1}{2\sigma_{1}^{2}}{\displaystyle \frac{i\mu{\displaystyle \left(\mu^{2}+\rho\mu^{2}+1\right)}}{\left(-i\mu+\xi\right)\left(\mu{\displaystyle \left(i\xi-\mu\right)}\rho+2\left(\xi-1\right)\left(i\mu+1\right)\right)}}\label{alphachi}\\
 & \xi\equiv\frac{d}{d_{1}}=\frac{z}{L}\nonumber 
\end{align}

where $\mu,\rho$ are defined in Eq.(\ref{P2b}). And (derivation
of $x_{c}$ is given in Appendix IV.3),

\begin{align}
 & x_{c}=\frac{{\displaystyle c_{s1}}s_{1}+{\displaystyle c_{s2}}s_{2}}{{\displaystyle \mu^{2}+\rho\mu^{2}+1}}\label{xc}\\
 & {\displaystyle c_{s1}}\equiv{\displaystyle \rho\mu^{2}-\left(i\mu+1\right)}\left(\xi-1\right)\nonumber \\
 & {\displaystyle c_{s2}}\equiv{\displaystyle \left(\mu-i\right)\left(\mu+i\xi\right)}\nonumber 
\end{align}

To simplify $C$ in Eq.(\ref{lnphipsi}), we apply the Gaussian integral
Eq.(\ref{gaussian_integral}), the formula of $P_{2b}$ Eq.(\ref{P2b}),
and $P_{2b}=\frac{1}{2}\int_{-\infty}^{\infty}dx\frac{\delta P_{2b}}{\delta\chi(x)}$
, and we have 

\begin{align}
 & \frac{1}{2}\int_{-\infty}^{\infty}dx\frac{\delta P_{2b}}{\delta\chi(x)}=\frac{1}{2}\int_{-\infty}^{\infty}dx\exp(\alpha_{\chi}\left(x-x_{c}\right)^{2}+C)+c.c.\nonumber \\
 & =\exp(C)\sqrt{\frac{\pi}{-\alpha_{\chi}}}=P_{2b}=\sqrt{\frac{\rho\mu^{2}}{{\displaystyle \mu^{2}+\rho\mu^{2}+1}}}\exp\left(-\frac{1}{2\sigma_{1}^{2}}\frac{\mu^{2}}{\mu^{2}+\rho\mu^{2}+1}(s_{1}-s_{2})^{2}\right)\label{dP2bdchi}\\
 & C=-\frac{1}{2\sigma_{1}^{2}}\frac{\mu^{2}}{\mu^{2}+\rho\mu^{2}+1}(s_{1}-s_{2})^{2}+\ln\sqrt{\frac{\rho\mu^{2}}{{\displaystyle \mu^{2}+\rho\mu^{2}+1}}}+\ln\frac{\sqrt{-\alpha_{\chi}}}{\sqrt{\pi}}\nonumber 
\end{align}

Compared with $C$ given in Eq.(\ref{lnphipsi}), we can separate
the terms quadratic in $s_{1},s_{2}$ and the logarithmic terms in
$C$ in Eq.(\ref{lnphipsi}). This separation simplifies the the calculation
to confirm that the complicated expression $C$ in Eq.(\ref{lnphipsi})
can be simplified significantly as the sum of the following 3 terms:

\begin{align}
 & \ln\frac{\sqrt{-\alpha_{\chi}}}{\sqrt{\pi}}+\ln\sqrt{\frac{\rho\mu^{2}}{{\displaystyle \mu^{2}+\rho\mu^{2}+1}}}=\frac{1}{2}\ln\frac{\beta_{2}}{i\pi}+\ln\left(\left(-\pi A\right)^{\frac{1}{2}}\right)+\frac{\ln\left(\frac{\sigma_{1}^{2}}{\Delta}\frac{\sigma_{1}^{2}}{\overline{\Delta_{1}}}\frac{1}{2\pi\sigma_{1}^{2}}\right)}{2}\label{C_separation}\\
 & -\frac{1}{2\sigma_{1}^{2}}\frac{\mu^{2}}{\mu^{2}+\rho\mu^{2}+1}(s_{1}-s_{2})^{2}=-\frac{\left(Ai\beta_{2}\left(\alpha_{f2}s_{2}+\overline{\alpha_{2b}}\overline{s_{2b}}\right)-s_{1}\frac{1}{4\Delta}\right)^{2}}{\alpha_{\chi}}\nonumber \\
 & -A\left(\alpha_{f2}s_{2}+\overline{\alpha_{2b}}\overline{s_{2b}}\right)^{2}-\frac{\left(s_{1}\right)^{2}}{4\Delta}+\alpha_{f2}s_{2}^{2}+\overline{\alpha_{2b}}\overline{s_{2b}}^{2}-\frac{(s_{1}-s_{2})^{2}}{4\left(\overline{\Delta_{1}}+\sigma_{2}^{2}\right)}\nonumber 
\end{align}

Substituting $C$ in Eq.(\ref{phipsi}) by the expression of $C$
in Eq.(\ref{dP2bdchi}), and using the expression of $P_{2b}$ in
Eq.(\ref{dP2bdchi}), the result is the analytical expression for
$\frac{\delta P_{2b}}{\delta\chi(x,z)},$

\begin{align}
 & \frac{\delta P_{2b}}{\delta\chi(x,z)}=P_{2b}\frac{\sqrt{-\alpha_{\chi}}}{\sqrt{\pi}}\exp(\alpha_{\chi}\left(x-x_{c}\right)^{2})+c.c.\label{dPdchi-1}
\end{align}

\subsection*{IV.2 Derivation of $\alpha_{\chi}$}

In Section IV.1 Eq.(\ref{lnphipsi}), we have

\begin{align}
 & \ln\phi_{b}(x)\psi(x)=\alpha_{\chi}\left(x-x_{c}\right)^{2}+C\label{lnphipsi-1}\\
 & \alpha_{\chi}=i\beta_{2}-A\left(i\beta_{2}\right)^{2}-\frac{1}{4\Delta}\nonumber \\
 & A\equiv\frac{1}{\alpha_{f2}+i\beta_{2}+\overline{\alpha_{2b}}}\nonumber 
\end{align}

$\alpha_{f2},\Delta,\Delta_{1}$ and $d_{1}=\frac{\lambda(z_{2}-z_{1})}{4\pi}=\frac{\lambda L}{4\pi}$
, $d=\frac{\lambda z}{4\pi}$ are given in section 3.2, $\beta_{2},d_{2}$
are defined right before Eq.(\ref{lnpsi2bs}). Substituting them into
$\alpha_{\chi}$, we have

\begin{align}
 & \beta_{2}=\frac{M}{2\hbar(t_{2}-t)}=\frac{k}{2(L-z)}\equiv\frac{1}{4d_{2}}=\frac{1}{4(d_{1}-d)}\nonumber \\
 & \Delta=\sigma_{1}^{2}+id\nonumber \\
 & \overline{\alpha_{2b}}=-\frac{1}{4\overline{\Delta_{1}}}-\frac{1}{4\sigma_{2}^{2}}=-\frac{1}{4\left(\sigma_{1}^{2}-id_{1}\right)}-\frac{1}{4\sigma_{2}^{2}}=-\frac{\sigma_{2}^{2}+\sigma_{1}^{2}-id_{1}}{4\left(\sigma_{1}^{2}-id_{1}\right)\sigma_{2}^{2}}\nonumber \\
 & \frac{1}{A}=-\frac{1}{4\sigma_{2}^{2}}+i\frac{1}{4(d_{1}-d)}-\frac{1}{4\left(\sigma_{1}^{2}-id_{1}\right)}-\frac{1}{4\sigma_{2}^{2}}=-\frac{1}{4}\left(\frac{2}{\sigma_{2}^{2}}-i\frac{1}{(d_{1}-d)}+\frac{1}{\left(\sigma_{1}^{2}-id_{1}\right)}\right)\label{apIII.1}\\
 & \alpha_{\chi}=i\frac{1}{4(d_{1}-d)}+\frac{\frac{1}{4}\frac{1}{(d_{1}-d)}}{-\frac{1}{4}\left(\frac{2}{\sigma_{2}^{2}}-i\frac{1}{(d_{1}-d)}+\frac{1}{\left(\sigma_{1}^{2}-id_{1}\right)}\right)}\frac{1}{4}\frac{1}{(d_{1}-d)}-\frac{1}{4\left(\sigma_{1}^{2}+id\right)}\nonumber 
\end{align}

After simplification we get

\begin{align}
 & \alpha_{\chi}=\frac{i}{2}\left(\frac{{\displaystyle d_{1}^{2}+\sigma_{1}^{4}+\sigma_{1}^{2}\sigma_{2}^{2}}}{\left(\sigma_{1}^{2}+id\right)\left[2(d_{1}-d)\left(\sigma_{1}^{2}-id_{1}\right)-\sigma_{2}^{2}\left(d+i\sigma_{1}^{2}\right)\right]}\right)\label{alphachi2}
\end{align}

Using the dimensionless parameters $\mu\equiv\frac{\sigma_{1}^{2}}{d_{1}},\rho\equiv\frac{\sigma_{2}^{2}}{\sigma_{1}^{2}},\xi\equiv\frac{d}{d_{1}}=\frac{z}{L}$
defined in Eq.(\ref{P2b}) in Section 3.3, this becomes

\begin{align}
 & \alpha_{\chi}=-\frac{1}{2\sigma_{1}^{2}}{\displaystyle \frac{i\mu{\displaystyle \left(\mu^{2}+\rho\mu^{2}+1\right)}}{\left(-i\mu+\xi\right)\left(\mu{\displaystyle \left(i\xi-\mu\right)}\rho+2\left(\xi-1\right)\left(i\mu+1\right)\right)}}\label{alphachi-1}
\end{align}

\subsection*{IV.3 Derivation of $x_{c}$}

In Section IV.1 Eq.(\ref{lnphipsi}) we have

\begin{equation}
x_{c}=\frac{\left(Ai\beta_{2}\left(\alpha_{f2}s_{2}+\overline{\alpha_{2b}}\overline{s_{2b}}\right)-s_{1}\frac{1}{4\Delta}\right)}{\alpha_{\chi}}\label{xc-1}
\end{equation}

where $A,\overline{\alpha_{2b}}$ are in Eq.(\ref{apIII.1}), $s_{2b}$
is in Eq.(\ref{parameters}), $\alpha_{f2}$ is given in Section 3,
substituting these and Eq.(\ref{alphachi2}) for $\alpha_{\chi}$,
we find $x_{c}=x_{cs1}+x_{cs2}$, each term is linear in $s_{1},s_{2}$
separately 

\begin{align}
 & {\displaystyle x_{cs1}=-\frac{s_{1}\left(id+\sigma_{1}^{2}\right)\left(\sigma_{2}^{2}\left(d+i\sigma_{1}^{2}\right)-2\left(-d_{1}+d\right)\left(id_{1}-\sigma_{1}^{2}\right)\right)\left(d_{1}^{2}-d_{1}d+id_{1}\sigma_{1}^{2}-id\sigma_{1}^{2}+\sigma_{1}^{2}\sigma_{2}^{2}\right)}{\left(d-i\sigma_{1}^{2}\right)\left(d_{1}^{2}+\sigma_{1}^{4}+\sigma_{1}^{2}\sigma_{2}^{2}\right)\left(2d_{1}^{2}-2d_{1}d+2id_{1}\sigma_{1}^{2}-2id\sigma_{1}^{2}-id\sigma_{2}^{2}+\sigma_{1}^{2}\sigma_{2}^{2}\right)}}\\
 & x_{cs2}={\displaystyle -\frac{s_{2}\left(d_{1}+i\sigma_{1}^{2}\right)\left(id+\sigma_{1}^{2}\right)\left(\sigma_{2}^{2}\left(d+i\sigma_{1}^{2}\right)-2\left(-d_{1}+d\right)\left(id_{1}-\sigma_{1}^{2}\right)\right)}{\left(d_{1}^{2}+\sigma_{1}^{4}+\sigma_{1}^{2}\sigma_{2}^{2}\right)\left(2d_{1}^{2}-2d_{1}d+2id_{1}\sigma_{1}^{2}-2id\sigma_{1}^{2}-id\sigma_{2}^{2}+\sigma_{1}^{2}\sigma_{2}^{2}\right)}}\nonumber 
\end{align}

The following common factors lead to cancellation

\begin{align}
 & 2d_{1}^{2}-2d_{1}d+2id_{1}\sigma_{1}^{2}-2id\sigma_{1}^{2}-id\sigma_{2}^{2}+\sigma_{1}^{2}\sigma_{2}^{2}\\
 & =-i\left(\sigma_{2}^{2}\left(d+i\sigma_{1}^{2}\right)-2\left(-d_{1}+d\right)\left(id_{1}-\sigma_{1}^{2}\right)\right)\nonumber 
\end{align}

Finally, we find $x_{c}$ expressed in terms of the scaled dimensionless
parameters $\mu,\rho$ in Eq.(\ref{P2b})

\begin{align}
 & x_{c}=\frac{{\displaystyle c_{s1}}s_{1}+{\displaystyle c_{s2}}s_{2}}{{\displaystyle \mu^{2}+\rho\mu^{2}+1}}\label{xc-3}\\
 & {\displaystyle c_{s1}}\equiv{\displaystyle \rho\mu^{2}-\left(i\mu+1\right)}\left(\xi-1\right)\nonumber \\
 & {\displaystyle c_{s2}}\equiv{\displaystyle \left(\mu-i\right)\left(\mu+i\xi\right)}\nonumber 
\end{align}

\end{document}